\documentclass[aps,11pt,superscriptaddress,longbibliography]{revtex4-1}

\usepackage[english]{babel}

\usepackage[letterpaper,top=2cm,bottom=2cm,left=3cm,right=3cm,marginparwidth=1.75cm]{geometry}

\usepackage{amsmath,amsfonts,amssymb}
\usepackage{graphicx}
\usepackage[colorlinks=true, allcolors=blue]{hyperref}
\usepackage{caption}
\usepackage{subcaption}
\usepackage{cleveref}
\usepackage{float}

\begin{document}
\title{Data-based approach for time-correlated closures of turbulence models}
\author{J. Domingues Lemos}%\email{juliah.domingues@gmail.com}
\affiliation{Universidade Federal do Rio de Janeiro - UFRJ}
\author{A. A. Mailybaev}
\affiliation{Instituto Nacional de Matemática Pura e Aplicada – IMPA}

\begin{abstract}
Developed turbulent motion of fluid still lacks an analytical description despite more than a century of active research. Nowadays phenomenological ideas are widely used in practical applications, such as small-scale closures for numerical simulations of turbulent flows. In the present work, we use a shell model of turbulence to construct a closure intended to have a solid theoretical background and to capture intrinsic probabilistic features of turbulence. Shell models of turbulence are dynamical deterministic systems used to model energy cascade and other key aspects of the Navier-Stokes  such as intermittency. We rescale the variables of the Sabra model in a way which leads to hidden symmetries and universal distributions. We then use such fine distributions to write closures, i.e., missing expressions for some of the Sabra variables. Our closures rely on approximating probability density functions using a Gaussian Mixture Model, which makes them probabilistic by nature and allows us to write time-correlated closures. We also provide a framework where other Machine Learning tools can be employed with reduced black-box aspects.
\end{abstract}
\maketitle

\section{Introduction}

A fully resolved simulation of a developed turbulent flow, which contemplates all scales of motion up to Kolmogorov's viscous scale, is unfeasible for many real world flows. In the context of the Navier-Stokes equation, dealing with closure problems is essentially handling the fact that there are more degrees of freedom than there are equations we are able to solve. Many approaches for closure problems in Navier-Stokes turbulence rely on phenomenologial predictions or averaging of fluctuations like, e.g., Reynolds-averaged Navier-Stokes equations and Large Eddy Simulations \cite{reynoldsaverage,les_1,pope2000turbulent}.  Some works, such as \cite{nudging} put effort into modelling error at small scales, while \cite{clos_eqfree,clos_eyink} go into more general approaches in the proposing of subgrid models. The overall idea is to introduce an effective small scale (cutoff), such that only the larger scales are resolved while smaller scales are modeled by the closure. 

Even though approximate schemes may be satisfactory for practical applications, little is clear about the fundamental question: Can one formulate a small-scale closure that provides an accurate description of the flow at all resolved scales? The major obstacle for answering this question in turbulence is the intermittency phenomenon \cite{frisch}, because intermittent velocity fields are not statistically self-similar at small scales. Thus, a proper closure must reproduce multi-fractal properties of the flow at and beyond a cutoff scale. Another important, though less evident condition, is related to the observations that solutions for the Navier-Stokes equations at very large Reynolds numbers are expected to evolve as a stochastic process triggered by a small-scale noise \cite{Lorenz69,leithkraich,ruelle,eyinknoise,massimum,Palmer}, i.e., the solutions are spontaneously stochastic \cite{Mailybaev_2016,Mailybaev_2017,RTbifbof,spont1}. This implies that a proper closure must be probabilistic, reproducing the intrinsic stochastic nature of small-scale turbulence. We conclude that the existence of accurate closure depends on the two phenomena: intermittency and spontaneous stochasticity, both still not well understood theoretically.

The similar problem can be formulated for shell models of turbulence~\cite{gled,goy,lucashell}, which are more tractable than the Navier-Stokes equation while preserving its key aspects, such as intermittency, energy cascade and spontaneous stochasticity. The attempt of formulating an accurate closure for a shell model was done in \cite{Biferale_2017}, where a probabilistic approach was combined with the use of Kolmogorov multipliers. The Kolmogorov multipleirs are ratios of velocity fluctuations at adjacent scales, which appear to have a universal (not intermittent) single-time statistics in the inertial interval of developed turbulence \cite{kolmogorov_1962,benzi,eyink2003}. As reported in \cite{Biferale_2017}, accuracy of the proposed closures was limited. One possible reason is that multi-time multiplier correlations are not taken into account in these closures. The multi-time multiplier correlations, however, are again intermittent, i.e., do not possess a universal statistics.

In this work we develop the formalism for probabilistic, time-correlated closures in the Sabra shell model of turbulence \cite{lvov}. This formalism describes the closures which are potentially accurate in both of the key properties: intermittency and stochasticity. To overcome the problem that velocity fluctuations have multi-fractal statistics across the inertial interval, we apply the space-time rescaling proposed in \cite{alexei,Mailybaev2022HiddenSI}. This rescaling recovers the hidden scale invariance of the multi-time and multi-scale statistics. As a consequence, the resulting closure becomes universal, scale independent, and can be properly conditioned to the resolved scales not only at a current time but also to a prehistory of computations. 

For a practical realization of our formalism, we employ the probability density estimation with Gaussian Mixture Models \cite{bishop}. This approach is convenient for data generation because the estimated density is computed as a weighted sum of Gaussian components, which can easily generate conditional samples of new, unseen instances of data. On the other hand, this method demonstrates limitations in the learning process, especially in cases where the density estimation translates into an optimization problem in a space of high dimension. Numerical tests were performed leading to a satisfactory but not systematically improving performance of the models. These tests indicate that more elaborate tools of Machine Learning must be combined with our formalism in order to achieve the accurate closure.

We discuss the Sabra shell model in Section \ref{sec:model} and the its rescaled version with the hidden scaling symmetry in Section \ref{sec:rescaling}. Section \ref{sec:clos} describes the closure problem. In Section \ref{sec:gmm} we discuss the process of density estimation and conditional sampling. We present numerical results  in Sections \ref{sec:results} for single-time closures and in Section \ref{sec:withcond} for closures with time 
conditioning. We summarize the results and discuss further perspectives in the discussion Section \ref{sec_disc}. The Appendix contains supplementary computations and tables.

\section{Shell model}\label{sec:model}

Shell models are infinite dimensional systems of ordinary differential equations which rely on a discretization of the phase space that considers a sequence of wavenumbers as a geometric progression: $|\mathbf{k}| = k_n = k_0 \lambda ^n$. Typically, $\lambda=2$ and $k_0 = 1$. The Sabra model~\cite{lvov}, which we are considering in this work, is given by the following set of equations
\begin{align}
    \frac{du_n}{dt} = i\big(k_{n+1}u_{n+2}u_{n+1}^* - \frac{1}{2}k_n u_{n+1}u_{n-1}^* + \frac{1}{2} k_{n-1}u_{n-1}u_{n-2}\big) - \nu k_n^2u_n + f_n. \label{eqn:sabra}
\end{align}
where $u_n \in \mathbb{C}$ is a complex-valued variable describing a velocity fluctuation at shell $n$, $\nu$ is the viscosity coefficient and $f_n$ is the forcing term. In this work, $f_1$ is a nonzero constant and $f_n = 0$ for $n\geq 2$. The Reynolds number is defined as $\mathrm{Re}=UL/\nu$, with the integral scale $L=1/k_0$ and the characteristic large-scale velocity $U = \sqrt{|f_1|/k_0}$. 

The system of equations (\ref{eqn:sabra}) mimics important properties of developed hydrodynamic turbulence such as the energy cascade to small scales and intermittency~\cite{lucashell,frisch}. The regime of fully developed turbulence corresponds to very large Reynolds numbers. In this case, one distinguishes a wide range of scales called the inertial interval, for which both forcing and dissipative terms can be neglected. Within the K41 approximation~\cite{frisch}, the inertial interval extends to the wavenumbers $1 \ll k/k_0 \ll \mathrm{Re}^{3/4}$; this estimate is rough, but only because of the anomalous corrections~\cite{frisch1991prediction,mailybaev2023hidden}. The forcing range corresponds to the wavenumbers $k_n/k_0 \sim 1$ at which the energy is produced by the work of large-scale forces. This energy is transported through the inertial interval until it dissipates at wavenumbers of the dissipation range, $k/k_0 \gtrsim \mathrm{Re}^{3/4}$.

In the context of shell models, the closure problem involves first setting a finite number of resolved scales $n = 1,\ldots,s$ to be computed by solving the equations of motion, while the remaining unresolved scales with $n > s$ are to be provided by the closure model. Considering equations \eqref{eqn:sabra} with $n=1,2, \dots, s$ leaves us with missing expressions for shell velocities $u_{s+1}$ and $u_{s+2}$, which are complex numbers. If the total amount of resolved scales is large enough to cover forcing range, inertial range and dissipation range, then one can set $u_{s+1} = u_{s+2} = 0$. In this situation, the simulation is considered to be fully resolved. The goal of the closure, however, is to set $s$ to a much larger scale (smaller wavenumber). Namely, we consider $k_s$ to be a wavenumber in the inertial interval. Then the closure consists of writing expressions for the shell speeds $u_{s+1}$ and $u_{s+2}$, which are the only variables missing in equations (\ref{eqn:sabra}) for $n = s-1$ and $s$.

\section{Rescaled variables and hidden symmetry}\label{sec:rescaling}

The major problem in defining closures for turbulence models is the intermittency. As a consequence, statistics collected in time are not universal with respect to the choice of $s$, since velocity fluctuations are getting more intermittent with the increase of $k_s$. Similarly, there is a lack of universality with respect to the forcing as well. It was noted, however, that universal statistics are recovered for single-time ratios of velocities~\cite{benzi,eyink2003,chen2003kolmogorov}, which are called Kolmogorov multipliers following the ideas of the K62 theory~\cite{kolmogorov_1962,frisch}. It was shown recently that this universality is a consequence of the hidden scaling symmetry of the equations of motion~\cite{alexei,Mailybaev2022HiddenSI}, which allows us to recover the full scale invariance in the inertial range by using a properly rescaled velocities and time. 

The rescaled variables are introduced by fixing a reference shell $m$ and setting the momentary temporal scale (turn-over time) for this shell as
\begin{gather}\label{eqn:Tm}
    T_m(t) = \left(k_0^2 U^2 + \sum_{n <m} k_n^2|u_n|^2\right)^{-1/2},
\end{gather}

\noindent where $U$ is the characteristic velocity defined above. The quantity $T_m$ is always positive and the sum in the parenthesis can be seen as accumulated enstrophy of shells up to (but not including) $m$. We then introduce the rescaled  time as
\begin{gather}
    \tau = \int_0^t \frac{dt'}{T_m(t')}. \label{eqn:taut}    
\end{gather}

The nonlinear change from original time $t$ to rescales time $\tau$ acts in such a way that the long stretches of time when shell velocities would be near zero are shrunk and short periods where velocities would rapidly vary are stretched out. This notion becomes clearer when we think of rescaled time increments $d\tau$ as a rescaling of original time increments $dt$ by a turnover time $T_m(t)$. Next, we introduce the relative shell number $N = n-m$ and define rescaled velocities $\mathcal{U}_N$ as
\begin{align}
    \mathcal{U}_N &= k_m T_m(t) u_{N+m}(t). \label{eqn:UN}
\end{align} 

We can rewrite the Sabra system from equation \eqref{eqn:sabra} in its rescaled version, which for the full (forced and viscous) system becomes
\begin{align}
    \frac{d\mathcal{U}_N}{d\tau} = i(k_{N+1}\mathcal{U}_{N+2}\mathcal{U}_{N+1}^* -\frac{1}{2}k_{N} \mathcal{U}_{N+1}\mathcal{U}_{N-1}^* + \frac{1}{2}k_{N-1}\mathcal{U}_{N-1}\mathcal{U}_{N-2}) \nonumber \\
    +\left( \xi_{total} - \nu k_{N+m}^2T_m \right) \mathcal{U}_{N} + T_m^2 k_m f_{N+m}, \label{eqn:system6}
\end{align}
where
\begin{gather} 
    \xi_{total} = \xi + \xi_{\nu} + \xi_f,  \label{eqn:dtmdt} \\
    \xi = \sum_{N<0} k_{N}^3 \operatorname{Im} \Big{(}  2 \mathcal{U}_{N}^*  \mathcal{U}_{N+1}^* \mathcal{U}_{N+2} - \frac{1}{2} \mathcal{U}_{N-1}^* \mathcal{U}_{N}^* \mathcal{U}_{N+1} - \frac{1}{4} \mathcal{U}_{N-1}^*  \mathcal{U}_{N}  \mathcal{U}_{N-2}^* \Big{)}, \label{eqn:xi}\\
    \xi_{\nu} = \nu T_m k_m^2 \sum_{N<0}  k_{N}^4 |\mathcal{U}_{N}|^2 , \label{eqn:xinu}\\
    \xi_f = - T_m^2 \sum_{N<0}k_{N+m} k_{N} \operatorname{Re}\Big{(} \mathcal{U}_{N}^* f_{N+m} \Big{)},  \label{eqn:xif} \\ 
    T_m = \frac{1}{k_0 U}\left( 1- \sum_{N<0}k_N^2|\mathcal{U}_N|^2 \right)^{1/2}. \label{eqn:Tm_tau}
\end{gather}
For the detailed derivation of Eqs.~(\ref{eqn:system6})--(\ref{eqn:Tm_tau}) see appendix \ref{app:proof}.

The hidden scale invariance \cite{alexei,mailybaev2022hidden} is the independence of the statistics of rescaled velocities $\mathcal{U}_N(\tau)$ in the inertial interval with respect to the choice of the reference shell $m$. Thus, rescaled variables restore the universality broken by the intermittency. This indicates that the presence of intermittent fluctuations, which prevented Sabra's statistics from being universal, has been encoded in the change of variables from \cref{eqn:taut,eqn:UN}. 

\begin{figure}
    \begin{subfigure}{.48\textwidth}
    \centering
        \subcaption{Absolute values}\label{fig:abs}\includegraphics[scale=0.48]{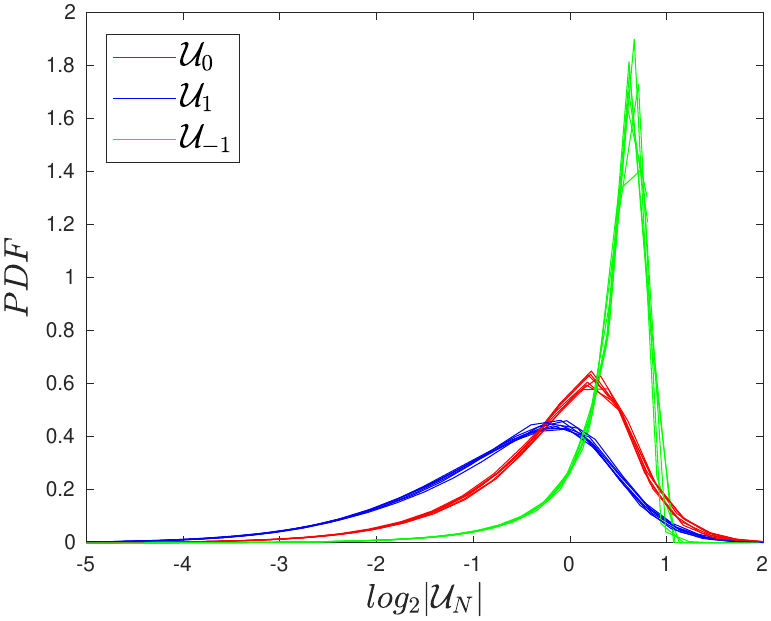}
    \end{subfigure}
    \begin{subfigure}{.48\textwidth}
    \centering
        \subcaption{Phases}\label{fig:phs}\includegraphics[scale=0.48]{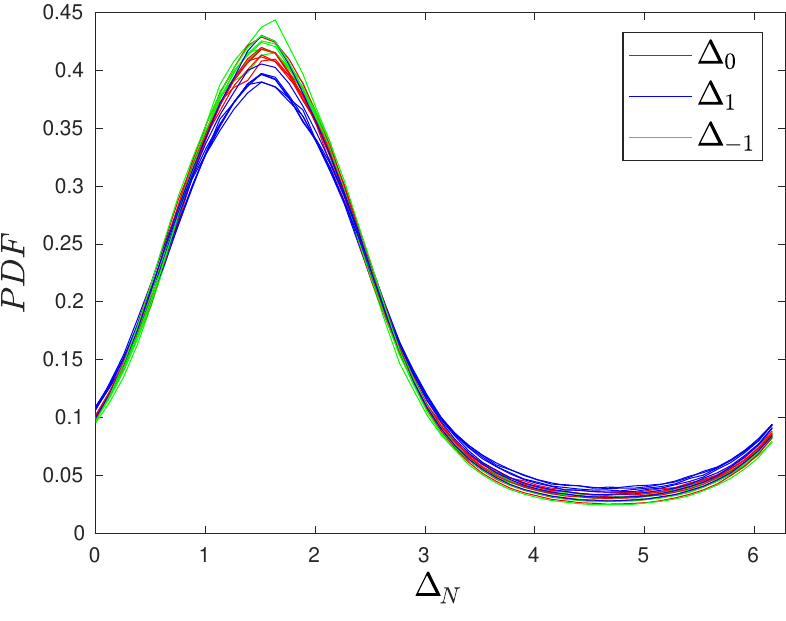}
    \end{subfigure}
    \caption{PDFs of absolute values and phases of $\mathcal{U}_{-1},\mathcal{U}_{0},\mathcal{U}_{1}$  $m=8,\dots,14$.}\label{fig:scaleinv}    
\end{figure}

To  observe hidden symmetry, we use the data from a simulation of fully resolved Sabra system with $30$ shells, while simultaneously solving for $\tau(t)$ using equation \eqref{eqn:taut}. In this simulation, $u_0 = u_{-1} = 0$, $\nu = 10^{-8}$, $f_1 = 1+i$ and initial condition is taken from a stationary state. The results are presented in Fig.~\ref{fig:abs} showing probability density functions (PDFs) of absolute values for $\mathcal{U}_N$ for $n = -1,0,1$. Colors indicate different values of $N$, while a number of the curves of the same color correspond to the reference shell varied in the inertial interval as $m = 8,\ldots,14$. The collapse of the curves of the same color verifies the hidden scale invariance.

Likewise, we can look at PDFs of the phases of rescaled variables. The simplest nontrivial (non-uniform) PDF corresponds to the phase differences of the form~\cite{benzi,eyink2003}
\begin{gather}
    \Delta_N = \arg(\mathcal{U}_N) - \arg(\mathcal{U}_{N-1}) - \arg(\mathcal{U}_{N-2}). \label{eqn:deltaa}
\end{gather}
PDFs for these phases are presented in Fig.~\ref{fig:phs}. Again, each of $\Delta_{-1}$, $\Delta_{0}$ and $\Delta_{1}$ is represented by a different color, and the curves of the same color collapse verifying the hidden scale invariance.

% Assuming we have access to one full simulation of original variables $u_n$, we can set a reference shell $m$ and compute rescaled variables $\mathcal{U}_N$. By choosing different values of $m$ we can look at different scales computing, for example, $\mathcal{U}_0$, which refers to the $m$-th scale. For Sabra, as first defined in \cite{kolmogorov_1962}, we can write the multipliers as
% \begin{gather}
%    z_n = w_n e^{i \delta_n}, \:\:\:\: w_n = \left| \frac{u_n}{u_{n-1}} \right|, \:\:\:\:  \delta_n = \arg(u_n) - \arg(u_{n-1}) - \arg(u_{n-2}), \label{eqn:d}
%\end{gather}

%\begin{figure}
%    \subfigure[ Absolute values]{\includegraphics[scale=0.48]{abs_8m14.pdf}}
%    \subfigure[ Phases]{\includegraphics[scale=0.48]{deltap_8m14_nn.pdf}}\label{fig:phs}
%    \caption{PDFs of absolute values and phases of $\mathcal{U}_{-1},\mathcal{U}_{0},\mathcal{U}_{1}$  $m=8,\dots,14$.}\label{fig:scaleinv}    
%\end{figure}

\section{Closure for a truncated rescaled system}\label{sec:clos}

In the context of the Sabra model, because the coupling of scales is local, writing a closure for a reduced model with $s$ scales translates into writing expressions for the closure variables $u_{s+1}$ and $u_{s+2}$. Because shell velocities are complex numbers, we need to compute absolute values and phases of each one. When $s$ is a scale of motion in the inertial range, it is unreasonable to set $u_{s+1} = u_{s+2} = 0$ because energy cascade is still at play and the viscous term is overcome by the nonlinear term and is, therefore, not strong enough to dissipate any energy.

In terms of rescaled variables, we need to first choose a reference shell. In this work we are committing to $m=s+1$, because this sets the new local temporal scale at shell $s+1$, which is one of the closure variables. With this setting, the closure variables for the rescaled system are $\mathcal{U}_0$ and $\mathcal{U}_1$; see equation \eqref{eqn:UN}.

There are two important aspects of the closures presented in this work. The first one is that they are probabilistic. Instead of fixing estimates for absolute values and phases of the closure variables, we are estimating the probability density functions of these quantities, like the ones presented in figures \ref{fig:abs} and \ref{fig:phs}, and sampling new instances of data from the density estimation. The new samples are used to evolve the reduced models in time.

The second one is the fact that the new instances of data for absolute values and phases can be sampled conditioned to the system's pre-history, which should take into account the fact that shell velocities are time-correlated. This raises the question of how long into the pre-history we should be looking. It is important to recall that the representation in rescaled variables $\mathcal{U}_N(\tau)$ has universal multi-time statistics \cite{alexei}, with the exceptional role played by the rescaling of time. This property is crucial for the formulation of pre-history dependent closures, which is the major advance of our formulation compared to the previous ideas based on Kolmogorov multipliers~\cite{benzi,eyink2003,Biferale_2017}.

\begin{figure}[]
\begin{subfigure}{.5\textwidth}
  \centering
  \subcaption{}
  \includegraphics[scale=0.6]{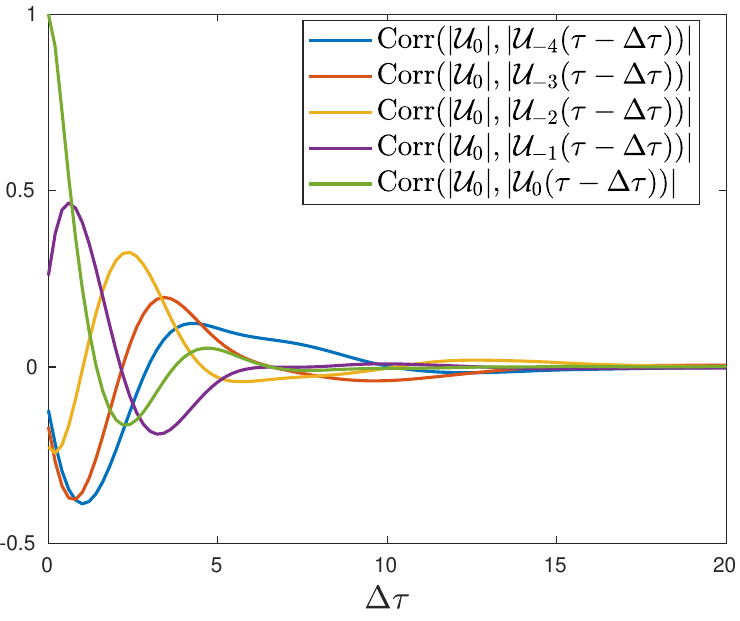}
  \label{fig:U01}
\end{subfigure}%
\begin{subfigure}{.5\textwidth}
  \centering
  \subcaption{}
  \includegraphics[scale=0.6]{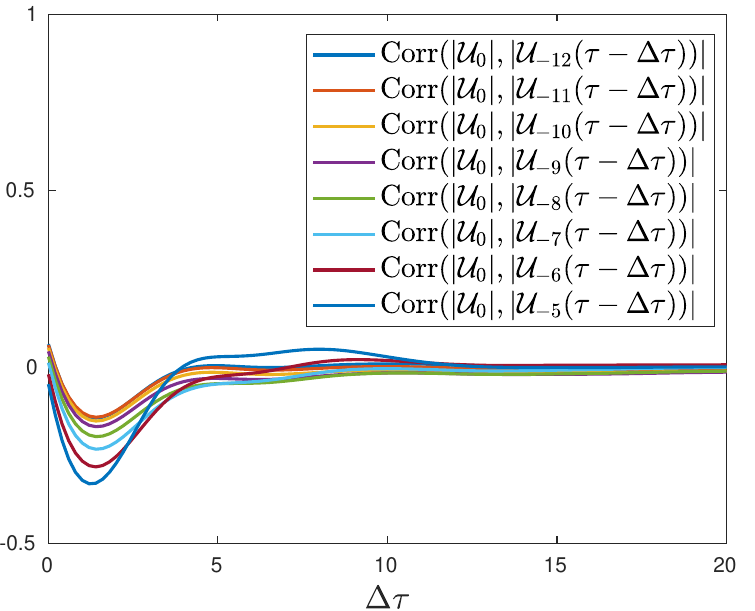}
  \label{fig:U02}
\end{subfigure}%

\begin{subfigure}{.5\textwidth}
  \centering
  \subcaption{}
  \includegraphics[scale=0.6]{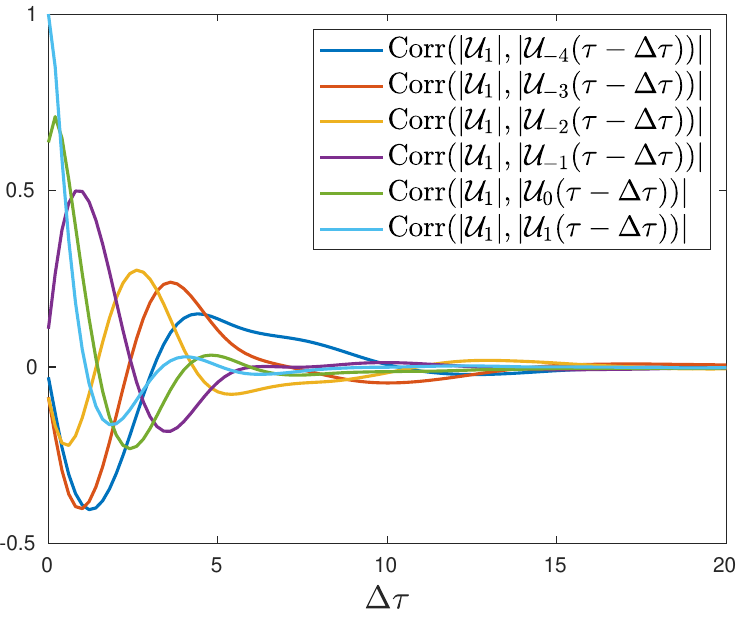}
  \label{fig:U11}
\end{subfigure}%
\begin{subfigure}{.5\textwidth}
  \centering
  \subcaption{}
  \includegraphics[scale=0.6]{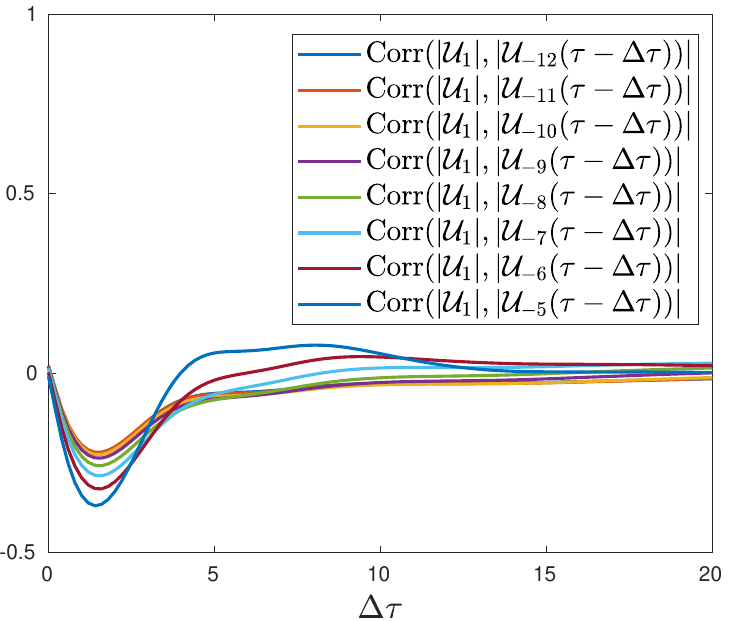}
  \label{fig:U12}
\end{subfigure}%
\caption{Figures \ref{fig:U01} and \ref{fig:U02} show correlation coefficients between $\mathcal{U}_0$ and other delayed shells, and figures \ref{fig:U11} and \ref{fig:U12} show correlation coefficients between $\mathcal{U}_1$ and other delayed shells for different values of $\Delta \tau$. Here, $s=12$ and $m=s+1$.}
\label{fig:corr}
\end{figure}

In figure \ref{fig:corr} we plot correlation coefficients between $\mathcal{U}_0$ and all the other previous shells for different time delays $\Delta \tau$, as well as correlation coefficients between $\mathcal{U}_1$ and all the other previous shells, also $\Delta \tau$-delayed. What we can see is that correlation decays as we progress into the past, as well as between different scales as they grow further apart in space. This indicates that the bulk of relevant information may be captured by conditioning the sampling of new instances of data to neighboring shells delayed some $\Delta \tau \in [0, 10]$ approximately. Such conditioning can be explicitly made in each time step, and, moreover, it is possible to condition to more than one value at a time.

In the rescaled reduced system when the shell $s$ corresponds to a scale of motion in the inertial range, the viscous terms are negligible. Recalling that we committed to $m=s+1$, the rescaled reduced system is given by
\begin{gather}
    \frac{d\mathcal{U}_N}{d\tau} = i(k_{N+1}\mathcal{U}_{N+2}\mathcal{U}_{N+1}^* -\frac{1}{2}k_{N} \mathcal{U}_{N+1}\mathcal{U}_{N-1}^* + \frac{1}{2}k_{N-1}\mathcal{U}_{N-1}\mathcal{U}_{N-2}) \nonumber \\
    +\left( \xi + \xi_f \right) \mathcal{U}_N + T_m^2 k_m f_{N+m}, \quad
    N = -s,\ldots,-1.
    \label{eqn:systemclosed}
\end{gather}
\noindent where $\xi$, $\xi_f$ and $T_m$ are given by equations \cref{eqn:xi,eqn:xif,eqn:Tm_tau}. These equations can be solved numerically for $N = -s,\ldots,-1$ as long as the closure provides values of $\mathcal{U}_0$ and $\mathcal{U}_1$ at each time step. In our closures, instead of the complex values $\mathcal{U}_0$ and $\mathcal{U}_1$, we find more convenient to use their absolute values and phases. Furthermore, instead of the phases of $\mathcal{U}_0$ and $\mathcal{U}_1$, one can reformulate the closure in terms of phase differences $\Delta_0$ and $\Delta_1$ given by Eq.~(\ref{eqn:deltaa})~\cite{Biferale_2017}.

\section{Gaussian Mixture Models}\label{sec:gmm}

Before we start writing closures, we need to discuss how the density estimation is performed. We are employing a well established method called Gaussian Mixture Models (GMM), which relies solely on data from the target distribution. In our case, the target distribution consists of absolute values of $\mathcal{U}_0$ and $\mathcal{U}_1$, as well as the associated multipliers' phases and any other values to which we wish to condition them. GMM works on the assumption that the distribution can be approximated as a weighted sum of Gaussian components. Given a probability density function $p$ and a positive amount of Gaussian components $K$, we want to find means $\mu_k$, variances $\Sigma_k$ and positive weights $\pi_k$ such that 

\begin{gather}
	p(\mathbf{z}) = \sum_{k=1}^{K} \pi_k \mathcal{N}(\mathbf{z}| \bm{\mu}_k,\bm{\Sigma}_k). \label{eqn:p(x)}
\end{gather}
 
\noindent Here, $\mathcal{N}(\mathbf{z}| \mu,\Sigma)$ denotes the normal distribution of mean $\mu$ and variance $\Sigma$, while $\mathbf{z} \in \mathbb{R}^d$ for integer $d \geq 1$. The parameters are computed as an iterative process using an Expectation-Maximization algorithm, which is based on maximum likelihood estimates. We initialize with random guesses. This process and algorithm have been detailed, for example, in \cite{bishop}.

The amount of Gaussian components $K$ is a parameter that has to be decided beforehand. Theoretical results guarantee that we can find a suitable approximation \cite{dempsterlaird}, given enough data and small overlapping of the Gaussian components \cite{emconv,samplesizePSUTKA}, which means there are systematical ways of choosing this parameter. However, some knowledge of what the data set looks like is surely beneficial to this process. 

What we saw in this work, though, was a relevant struggle on the method's performance when dealing with high dimensional data. This relates closely to the fact that Euclidean distances between data points increase exponentially with the dimension, which turns the data set into a sparse one \cite{bellman,curse}. This can be mitigated by increasing the amount of data samples fed to the density estimator, but only up to a certain point, limited by finite data and RAM memory.

% \subsection{Sampling}

The choice of GMM for the density estimation relates heavily to the sampling process. To evolve reduced models in time we need new instances of data from the densities being approximated, and Gaussian generators are widely available. Once we are in possession of an estimate, to obtain a new datum we can use the weights $\pi_k$ to choose one of the Gaussian components and then use any in-built Gaussian generator with mean $\mu_k$ and variance $\Sigma_k$ to generate a new sample.

If we want to sample conditionally, once the Gaussian component is selected, it is enough to recalculate the mean and the variance of the selected component since any slice of a Gaussian distribution is another Gaussian distribution \cite{condmeanvar}. Given a concatenated random variable $Y = [y_1, y_2]$, $Y \sim \mathcal{N}(\mu,\Sigma)$, we write
\begin{gather}
    \mu = [\mu_1, \mu_2], \quad
    \Sigma = 
    \begin{bmatrix}
        \Sigma_{11} & \Sigma_{12}\\
        \Sigma_{21} & \Sigma_{22}
    \end{bmatrix}.
\end{gather}
with compatible dimensions. Then, $p(y_1|y_2) = \mathcal{N}(\bar\mu,\bar\Sigma)$
with~\cite{condmeanvar}
\begin{gather}
    \bar\mu = \mu_1 +\Sigma_{12}\Sigma_{22}^{-1}(y_2-\mu_2), \quad
    \bar\Sigma = \Sigma_{11} - \Sigma_{12}\Sigma_{22}^{-1}\Sigma_{21}.
    \label{eqn:covbarra}
\end{gather}

This process for generating a new data sample happens at each time step in the process of evolving the reduced model in time. We intend on sampling absolute values and multiplier's phases for $\mathcal{U}_0$ and $\mathcal{U}_1$, as presented in figures \ref{fig:abs} and \ref{fig:phs}, while conditioning them to the pre-history of shell velocities. Once sampled, these values are used in equation \eqref{eqn:systemclosed} in order to close the system.

\section{Single-time closures}\label{sec:results}

As we described in section \ref{sec:rescaling}, the universal closure of the shell model can be formulated in terms of rescaled variables. These variables define scale invariant statistics for closure variables $\mathcal{U}_0$ and $\mathcal{U}_1$, namely PDFs of module and phase of each one. Now we apply the method described in Section \ref{sec:gmm} to estimate such densities and generate new data under the same distributions to evolve reduced models in time. The reduced models we simulate were discussed in section \ref{sec:clos} and are given by equation \eqref{eqn:systemclosed}.

Below we formulate different closures given by expressions for $\mathcal{U}_0$ and $\mathcal{U}_1$. Then, we evolve the reduced models and compare statistics to the baseline fully resolved Sabra model. For the comparison, we look at PDFs of energy flux $\Pi_n$ from shell $n$ to $n+1$, which is given by 
\begin{gather}
    \Pi_n  = \operatorname{Im}(k_{n+1}u_{n+2}u_{n+1}^*u_n^* +\frac{1}{2}k_{n}u_{n+1}u_{n}^*u_{n-1}^* ), \label{eqn:fluxshell}
\end{gather}
as well as moments of order $p$ defined as
\begin{gather}
    S_p(k_n) =\langle |u_n|^ p \rangle, \label{eqn:moment}
\end{gather}
where the averaging is with respect to time. Also, we compare normalized PDFs of real parts of velocities. The summary of all closures, including other choices for slightly different closures not reported in the main text, is given in Appendix \ref{app:all}. In this section, we consider a simpler class of single-time closures, while the more elaborated closures with multi-time conditioning are studied in the next section.

It is important to note that, while the reduced models and closures are written for rescaled variables, the simulations are all run for the original velocities in rescaled time. In each time step we switch to rescaled variables in order to compute the closure variables, and then switch back to original variables to evolve it in time. This was a friendlier alternative to computing rescaled variables in rescaled time, both numerically and for comparison purposes, and it is done noting that
\begin{gather}\label{eqn:dudtau}
    \frac{du_n}{d\tau} = \frac{du_n}{dt} \frac{dt}{d\tau} = \frac{du_n}{dt}T_m.
\end{gather}
Each simulation was run until the rescaled time $\tau = 30000$ with time step $10^{-4}$ using an Adams-Bashforth scheme \cite{adb}. 

% In this expression, as we computed in chapter \ref{ch:rescaled},
 
% \begin{gather} 
%     \xi = \sum_{N<0} k_{N}^3 \operatorname{Im} \Big{(}  2 \mathcal{U}_{N}^*  \mathcal{U}_{N+1}^* \mathcal{U}_{N+2} - \frac{1}{2} \mathcal{U}_{N-1}^* \mathcal{U}_{N}^* \mathcal{U}_{N+1} - \frac{1}{4} \mathcal{U}_{N-1}^*  \mathcal{U}_{N}  \mathcal{U}_{N-2}^* \Big{)}, \tag{\ref{eqn:xi}}\\
%     \xi_f = - T_m^2 \sum_{N<0}k_{N+m} k_{N} \operatorname{Re}\Big{(} \mathcal{U}_{N}^* f_{N+m} \Big{)}  \tag{\ref{eqn:xif}}, \\ 
%     T_m = \frac{1}{k_0 U}\left( 1- \sum_{N<0}k_N^2|\mathcal{U}_N|^2 \right)^{1/2}. \tag{\ref{eqn:Tm_tau}}
% \end{gather}

In the simplest closure, called MJoint, we consider the absolute values of the rescaled variables $|\mathcal{U}_0|$ and $|\mathcal{U}_1|$ only, while their phases will be fixed by the maximum energy flux condition.
For determining distribution of these absolute values, we feed to the GMM a data set consisting of pairs  $(\log_2|\mathcal{U}_0|,\log_2|\mathcal{U}_1|) = (z_0,z_1)$ using three two-dimensional Gaussian components. Then, once we have an estimation $g$ for the density of the random vector $\mathbf{z} = (z_0,z_1)$, for each time step we can sample one $\mathbf{z}$ and use it to evolve the reduced model in time. This closure is written as follows.
\begin{gather} 
\mathcal{U}_{0} = 2^{z_0} e^{i(\frac{\pi}{2} + \alpha_{-1} + \alpha_{-2})}, \label{eqn:dist20}\\
\mathcal{U}_{1} = 2^{z_1} e^{i(\frac{\pi}{2} + \alpha_{0} + \alpha_{-1})}, \label{eqn:dist21}\\
\mathbf{z} = (z_0,z_1) \sim g(\mathbf{z}). \label{eqn:dist2}
\end{gather} 
where $\alpha_N = \arg (\mathcal{U}_N)$. This closure only takes into account the modules and fixes multipliers phases in $\pi/2$, which is the value of highest dissipation \cite{Biferale_2017}.

An analogous but more elaborate closure, called PJoint, includes both absolute values and phases of the rescaled variables $|\mathcal{U}_0|$ and $|\mathcal{U}_1|$. In this case, we feed the density estimator data of the form $(\log_2|\mathcal{U}_0|, \Delta_0, \log_2|\mathcal{U}_1|,\Delta_1) = (z_0,z_1,z_2,z_3)$, using six Gaussian components to compute the approximation. Then the closure is formulated as
\begin{gather}
    \mathcal{U}_{0} = 2^{z_0} e^{i(z_1 + \alpha_{-1} + \alpha_{-2})}, \label{eqn:both0}\\
    \mathcal{U}_{1} = 2^{z_2}e^{i(z_3 + \alpha_{0} + \alpha_{-1})}, \\
    \mathbf{z} = (z_0,z_1,z_2,z_3) \sim g(\mathbf{z}). \label{eqn:both}
\end{gather}

\begin{figure}[]
\begin{subfigure}{.333\textwidth}
  \centering
  \subcaption{$n=12$}
  \includegraphics[scale=0.35]{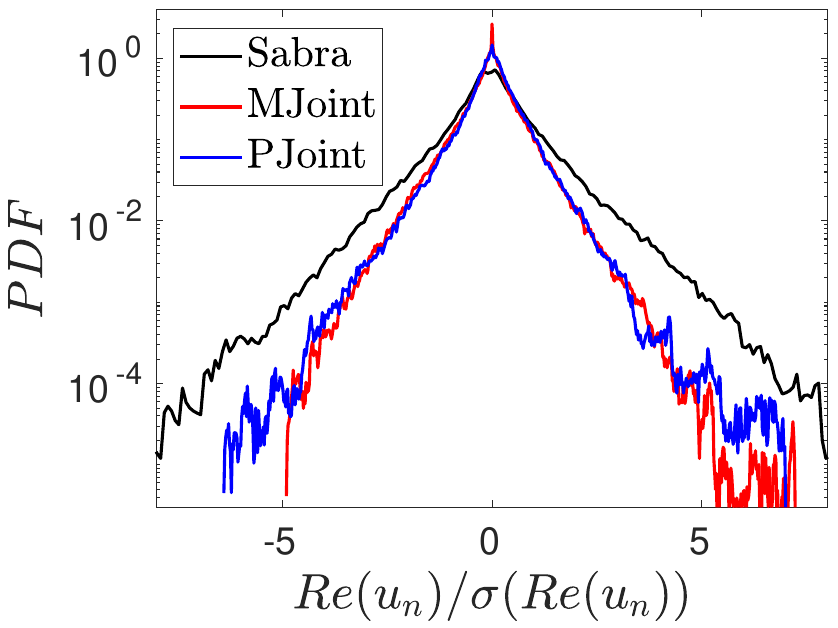}
  \label{fig:re12}
\end{subfigure}%
\begin{subfigure}{.333\textwidth}
  \centering
  \subcaption{$n=13$}
  \includegraphics[scale=0.35]{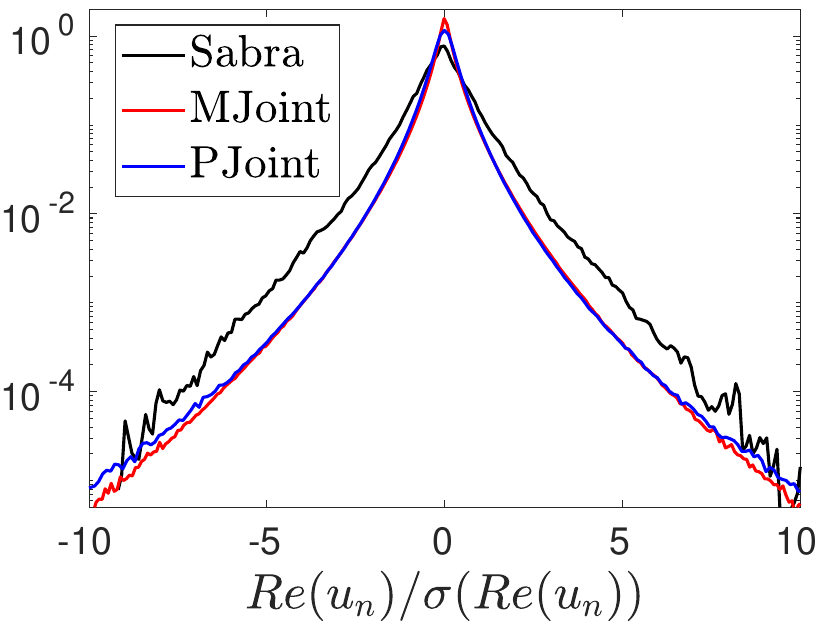}
  \label{fig:re13}
\end{subfigure}%
\begin{subfigure}{.333\textwidth}
  \centering
  \subcaption{$n=14$}
  \includegraphics[scale=0.35]{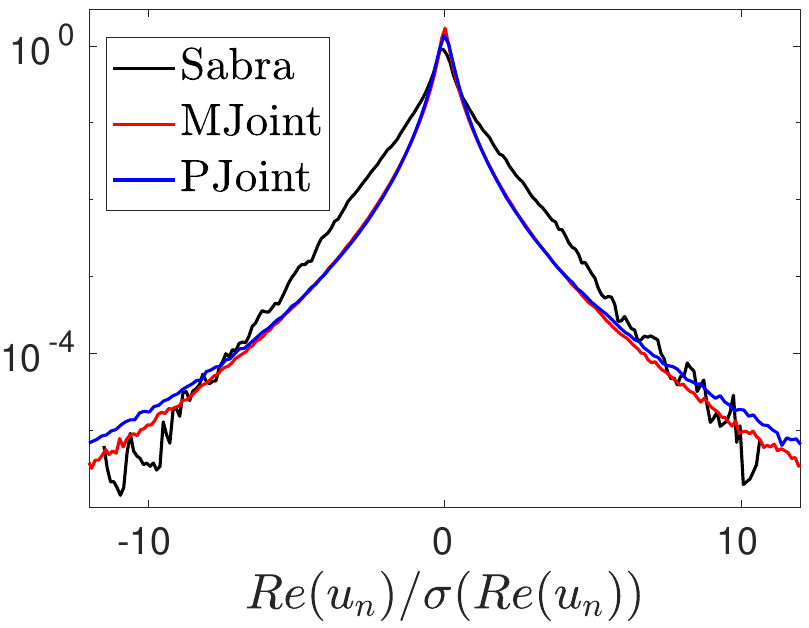}
  \label{fig:re14}
\end{subfigure}%

\begin{subfigure}{.5\textwidth}
  \centering
  \subcaption{MJoint}
  \includegraphics[scale=0.56]{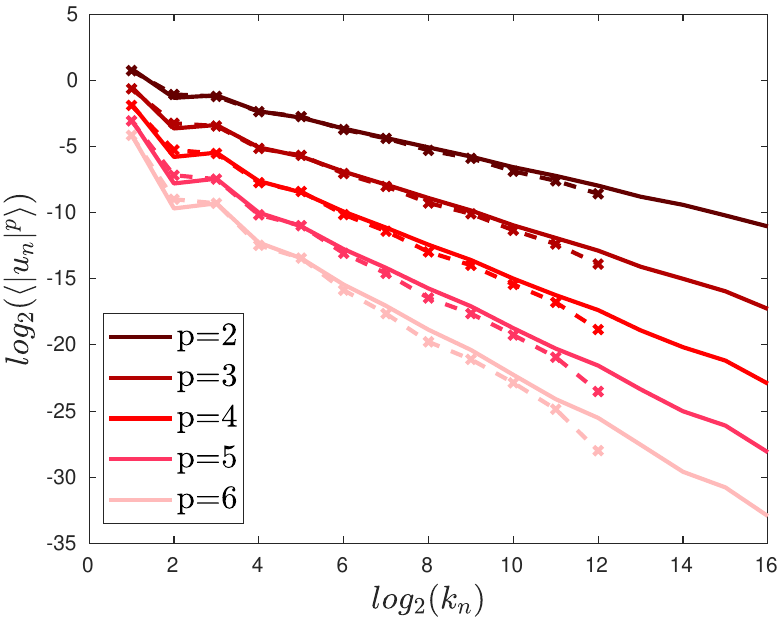}
  \label{fig:mjoint}
\end{subfigure}%
\begin{subfigure}{.5\textwidth}
  \centering
  \subcaption{PJoint}
  \includegraphics[scale=0.56]{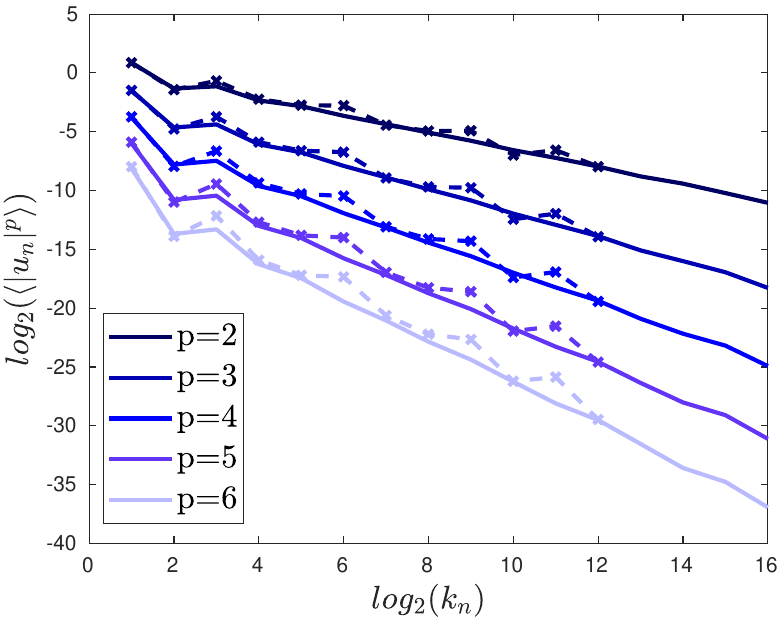}
  \label{fig:pjoint}
\end{subfigure}%

\begin{subfigure}{.333\textwidth}
  \centering
  \subcaption{$n=9$}
  \includegraphics[scale=0.34]{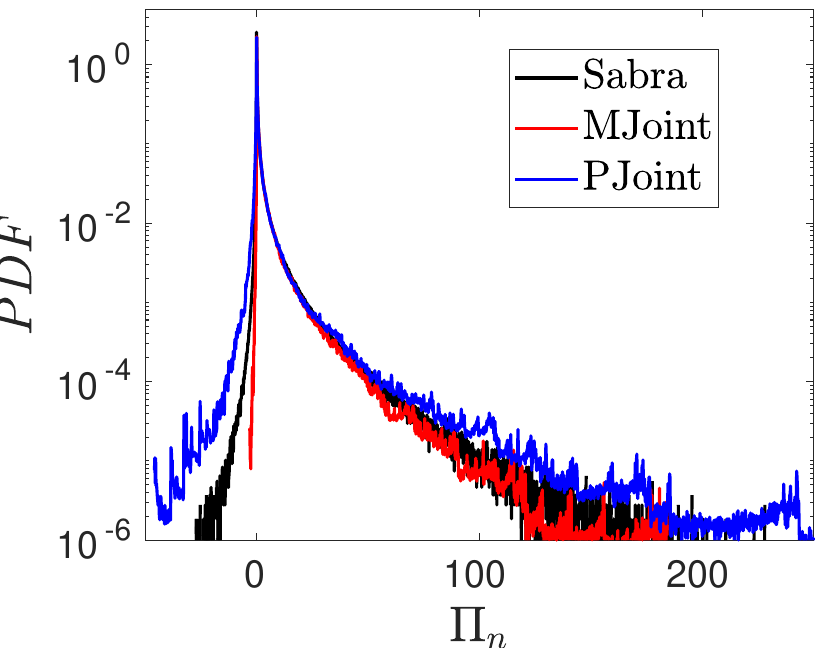}
  \label{fig:flx9}
\end{subfigure}%
\begin{subfigure}{.333\textwidth}
  \centering
  \subcaption{$n=11$}
  \includegraphics[scale=0.35]{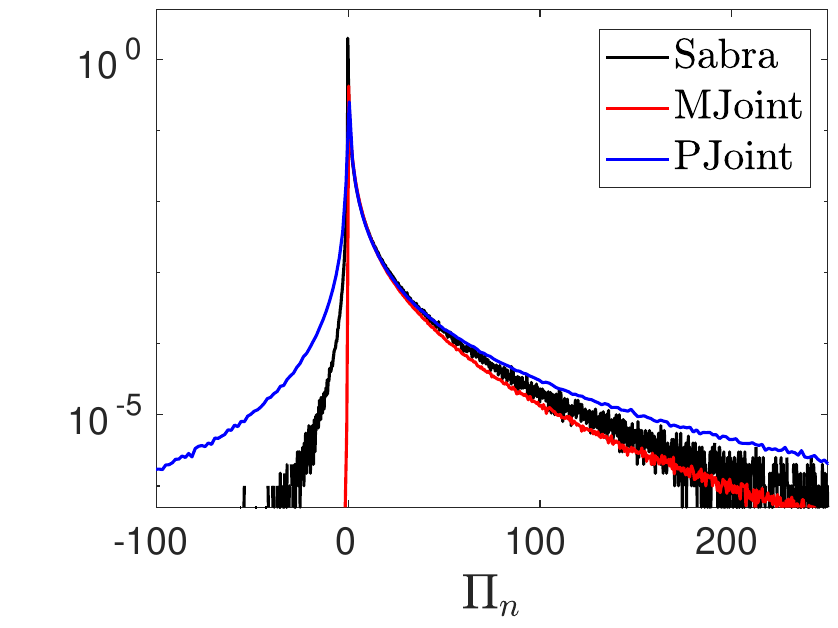}
  \label{fig:flx11}
\end{subfigure}%
\begin{subfigure}{.333\textwidth}
  \centering
  \subcaption{$n=12$}
  \includegraphics[scale=0.35]{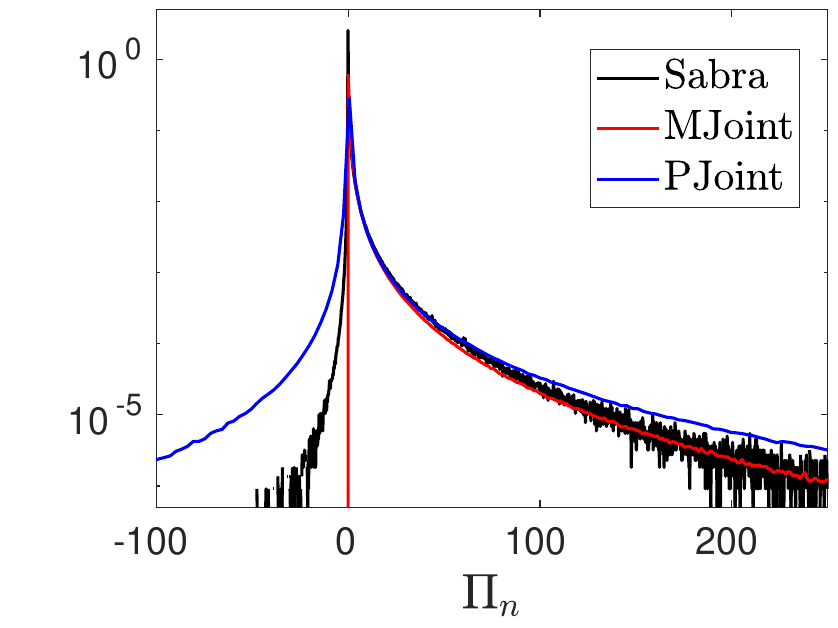}
  \label{fig:flx12}
\end{subfigure}%
\caption{Cut-off shell is $s=12$. Figures \ref{fig:re12}---\ref{fig:re14} present normalized PDFs of real parts between the closure with modules modelling (MJoint) and the closure with modules and phases modelling (PJoint). Figures \ref{fig:mjoint} and \ref{fig:pjoint} show moments of orders 2 up to 6 and figures \ref{fig:flx9}---\ref{fig:flx12} show energy flux PDFs.}
\label{fig:mpjoint}
\end{figure}

For a cut-off shell of $s=12$ we can see in Figs. \ref{fig:re12}---\ref{fig:re14} the PDFs for real parts of velocities in shells $12$ through $14$. Figure \ref{fig:mjoint} shows moments of order 2 up to 6 of the closure with module modelling, given by equations \eqref{eqn:dist20}---\eqref{eqn:dist2}. Figure \ref{fig:pjoint} shows moments of order 2 up to 6 for the closure with modules and phases, given by equations \eqref{eqn:both0}---\eqref{eqn:both}. In figures \ref{fig:flx9}---\ref{fig:flx12} we see energy flux PDFs for shells 9 through 12.

Both closures reproduce statistics to some degree of fidelity, especially moments. It is important to note that the closure without any phase modelling does not present negative flux of energy in the cut-off shell, which is due to the fixing of the phases at a value that strictly dissipates energy. On the other hand, including phases in the modelling introduces negative flux with higher probability than required. 

\section{Closures with time conditioning}\label{sec:withcond}

Using the conditional sampling method we introduced in section \ref{sec:gmm}, we can write closures that take the pre-history of the reduced model into account. Here the use of rescaled variables is crucial, because only in these variables one can recover the universal multi-time statistics. To that end, we train the density estimator on a data set containing the modules and phases of the closure variables, which are to be sampled, and all the variables in which we wish to condition. Both closures in this section use six Gaussian components in the density estimation.

Given that the correlation between shell velocities decays in time and in space, the first closure in this section will sample only modules of the closure variables conditioned to the modules of the last three shells of the reduced model. These are the ones representing the most significant correlation in space. Different values of delays $\Delta \tau$ yield different results in the conditioning process due to the fact that when the delay is either too small or too long, the correlation is either too high or essentially nonexistent. Recalling figure \ref{fig:corr}, we remark that different values of $\Delta \tau $ were initially tested to select a suitable delay and are reported in \cite{memyselfandi}. The delay selected for our simulations below is $\Delta \tau = 2.4$. Then the closure, called 3Most, can be written as
\begin{gather} 
\mathcal{U}_{0} = 2^{z_0} e^{i(\frac{\pi}{2} + \alpha_{-1} + \alpha_{-2})}, \label{eqn:3most0}\\
\mathcal{U}_{1} = 2^{z_1} e^{i(\frac{\pi}{2} + \alpha_{0} + \alpha_{-1})}, \label{eqn:3most1}\\
\mathbf{z} = (z_0,z_1) \sim g(\mathbf{z}| \log_2|\mathcal{U}_{-3}(\tau - \Delta \tau)|,\log_2|\mathcal{U}_{-2}(\tau - \Delta \tau)|,\log_2|\mathcal{U}_{-1}(\tau - \Delta \tau)|), \label{eqn:dist3most}
\end{gather} 
The last expression signifies that the closed variables are selected randomly with a given distribution, which is conditioned on the three larger-scale absolute velocities taken at time $t-\Delta \tau$.

Another type of conditioning can be performed by taking into account information of the closure variables themselves, allowing them to evolve conditioned to their own pre-history. In this self conditioning closure we are also sampling (and conditioning to) multipliers phases and the delay used in the conditioning is also $\Delta \tau = 2.4$. We can write this closure, called Self, as
\begin{gather}
    \mathcal{U}_{0} = 2^{z_0} e^{i(z_1 + \alpha_{-1} + \alpha_{-2})}, \label{eqn:s1s2ems1s20}\\
    \mathcal{U}_{1} = 2^{z_2}e^{i(z_3 + \alpha_{0} + \alpha_{-1})}, \\
    \mathbf{z} = (z_0,z_1,z_2,z_3) \sim g(\mathbf{z}|\log_2|\mathcal{U}_{0}(\tau - \Delta \tau)|,  \Delta_{0},\log_2|\mathcal{U}_{1}(\tau - \Delta \tau)|,  \Delta_{1}). \label{eqn:s1s2ems1s2}
\end{gather}
It is important to mention that, for both closures, the expressions above make sense for $\tau \geq \Delta \tau$. In the initial stage of simulation, when $0 < \tau < \Delta \tau$, one can employ some deterministic closure like the one described in \cite{Biferale_2017}.

\begin{figure}[t]
\begin{subfigure}{.333\textwidth}
  \centering
  \subcaption{$n=12$}
  \includegraphics[scale=0.35]{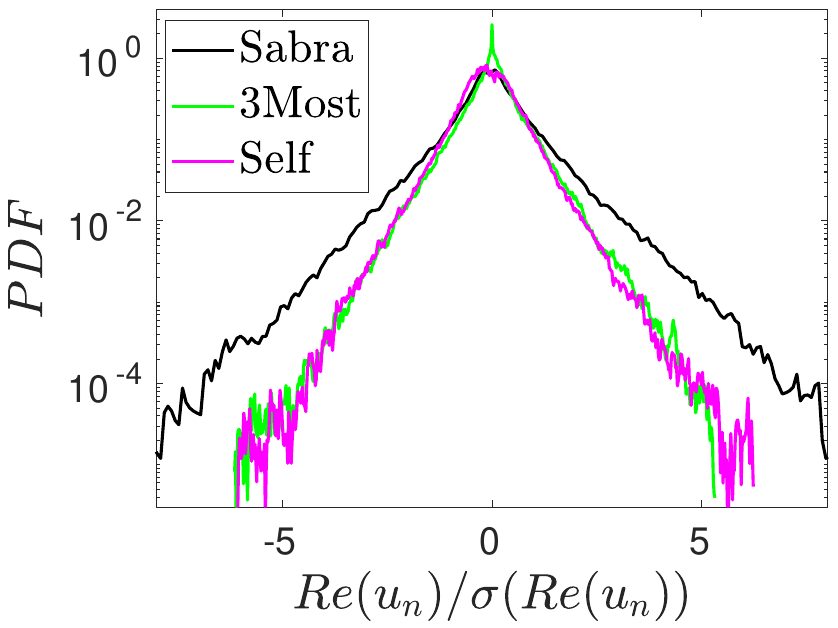}
  \label{fig:re123es}
\end{subfigure}%
\begin{subfigure}{.333\textwidth}
  \centering
  \subcaption{$n=13$}
  \includegraphics[scale=0.35]{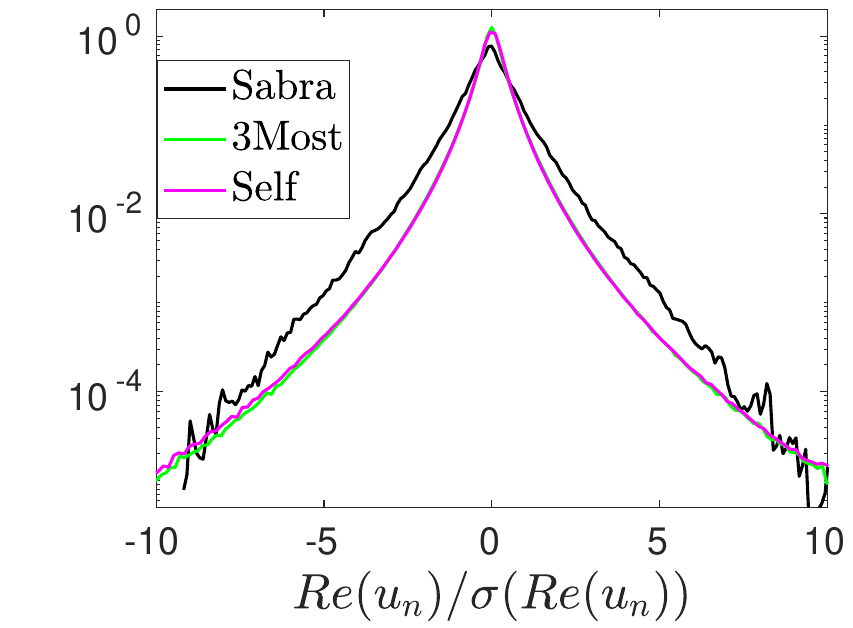}
  \label{fig:re133es}
\end{subfigure}%
\begin{subfigure}{.333\textwidth}
  \centering
  \subcaption{$n=14$}
  \includegraphics[scale=0.35]{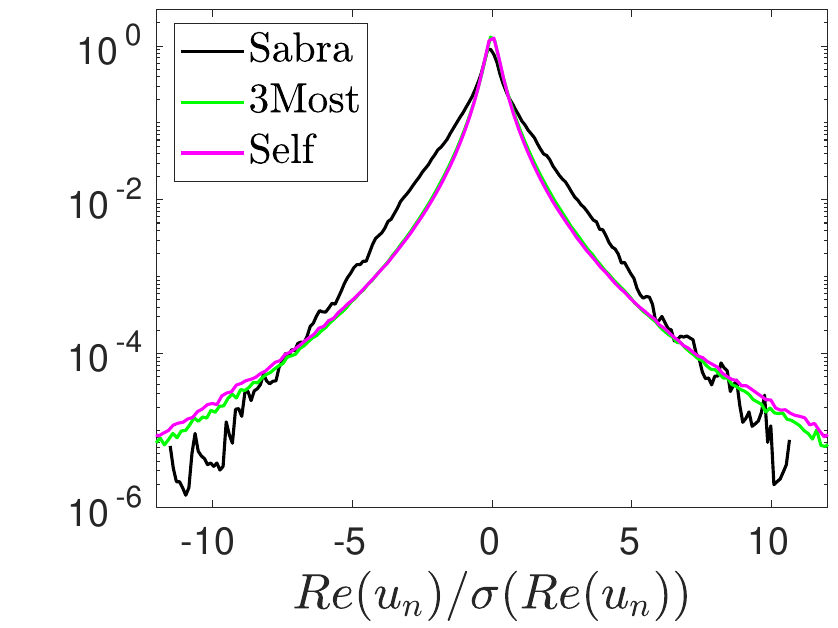}
  \label{fig:re143es}
\end{subfigure}%

\begin{subfigure}{.5\textwidth}
  \centering
  \subcaption{Self}
  \includegraphics[scale=0.56]{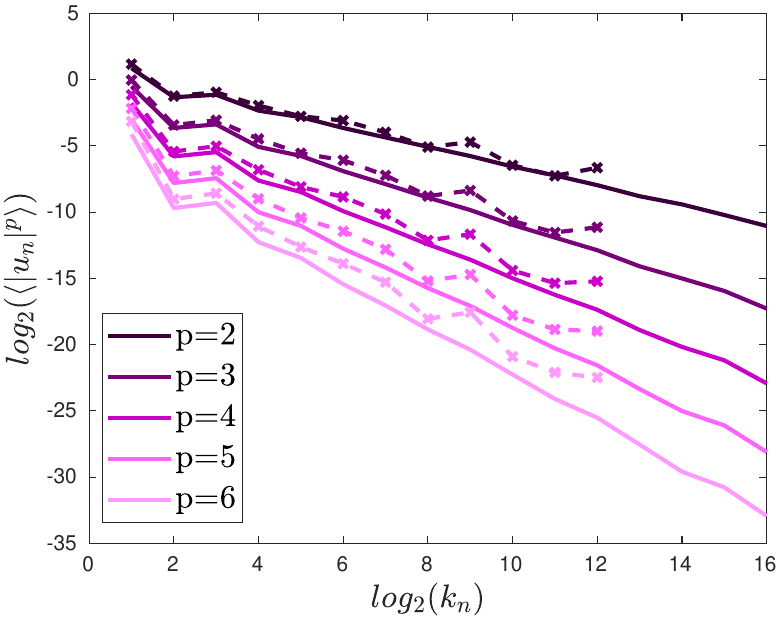}
  \label{fig:self12}
\end{subfigure}%
\begin{subfigure}{.5\textwidth}
  \centering
  \subcaption{3Most}
  \includegraphics[scale=0.56]{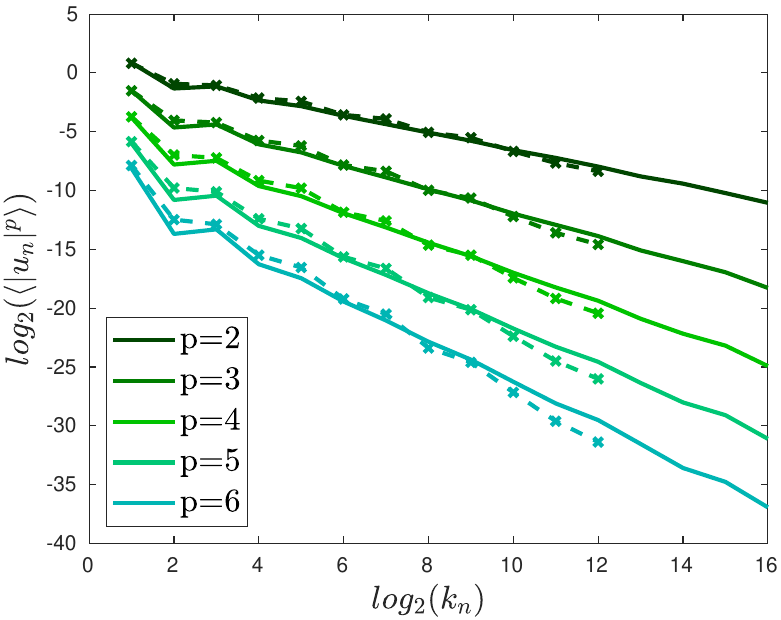}
  \label{fig:3cl12}
\end{subfigure}%

\begin{subfigure}{.3333\textwidth}
  \centering
  \subcaption{$n=9$}
  \includegraphics[scale=0.33]{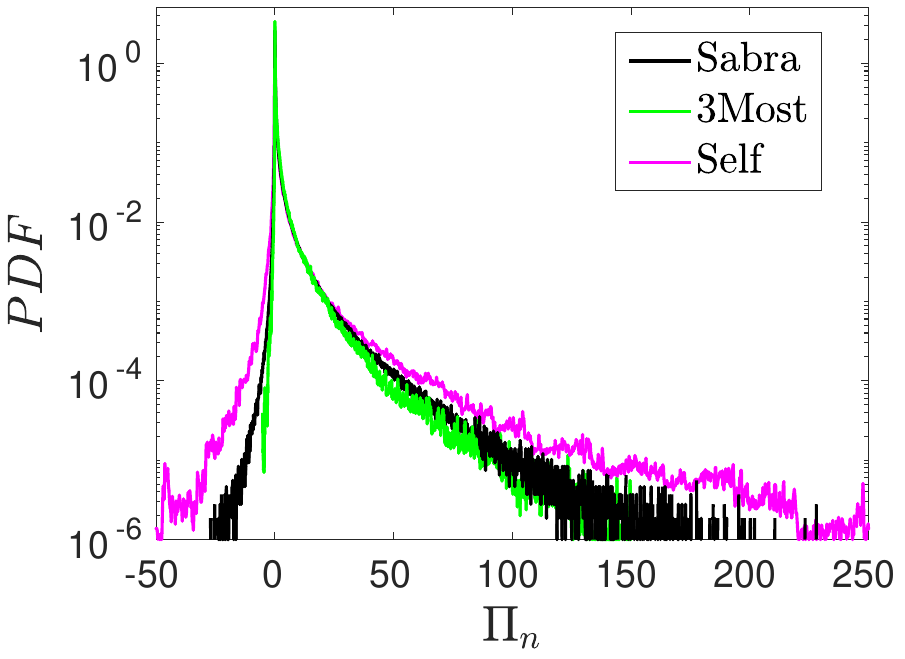}
  \label{fig:flx93es}
\end{subfigure}%
\begin{subfigure}{.3333\textwidth}
  \centering
  \subcaption{$n=11$}
  \includegraphics[scale=0.35]{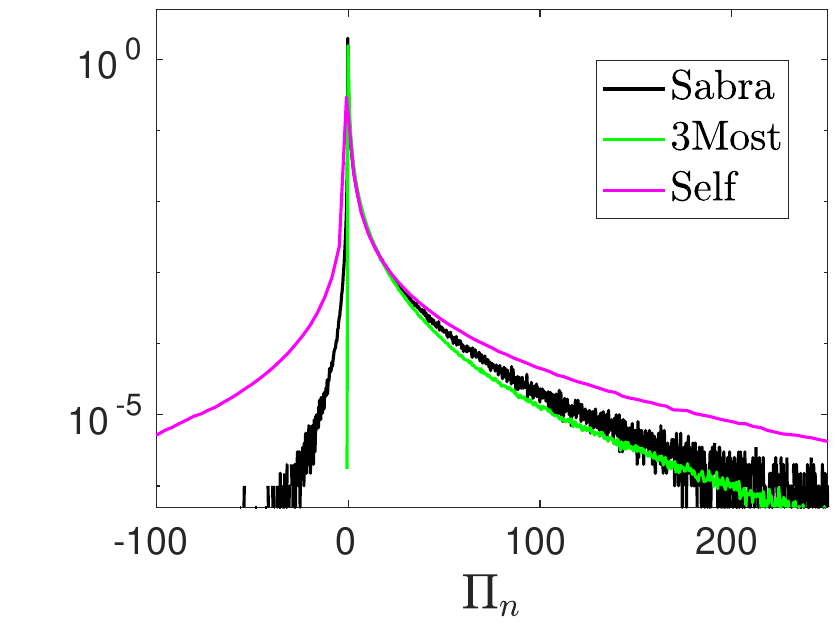}
  \label{fig:flx113es}
\end{subfigure}%
\begin{subfigure}{.3333\textwidth}
  \centering
  \subcaption{$n=12$}
  \includegraphics[scale=0.35]{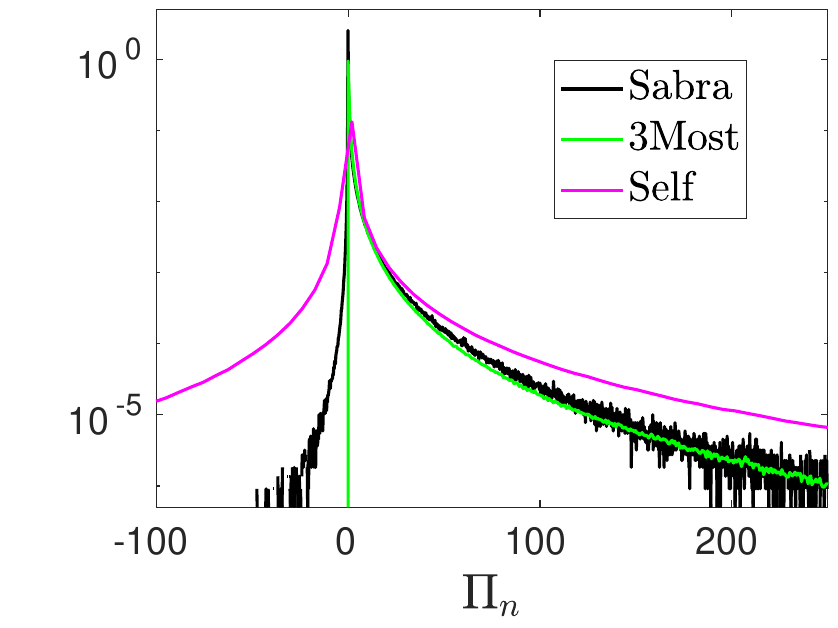}
  \label{fig:flx123es}
\end{subfigure}%
\caption{Cut-off shell is $s=12$. Figures \ref{fig:re123es}---\ref{fig:re143es} present normalized PDFs of real parts for the closure with time conditioning to the three closes shells (3Most) and the closure with time conditioning to the closure variables (Self). Figures \ref{fig:self12} and \ref{fig:3cl12} show moments of orders 2 up to 6 and figures \ref{fig:flx93es}---\ref{fig:flx123es} show energy flux PDFs.}
\label{fig:3eself}
\end{figure}

It is relevant to note that even with time conditioning, omitting the modelling of phases still produces drastic differences in statistics. The lack of backscattering in energy flux PDFs is, again, characteristic of the fixing phases to strictly dissipating values for the model \eqref{eqn:3most0}--\eqref{eqn:dist3most}. On the other hand, the inclusion of phases in the model \eqref{eqn:s1s2ems1s20}--\eqref{eqn:s1s2ems1s2} introduces negative energy flux events, but in an exaggerated manner.

\begin{figure}[t]
\begin{subfigure}{.333333\textwidth}
  \centering
  \subcaption{$n=9$}
  \includegraphics[scale=0.35]{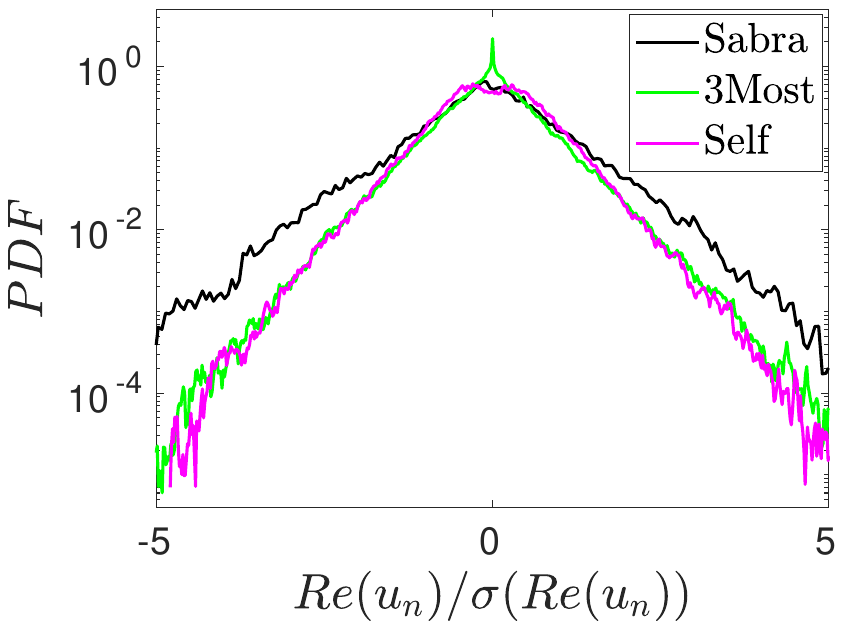}
  \label{fig:re129es}
\end{subfigure}%
\begin{subfigure}{.333333\textwidth}
  \centering
  \subcaption{$n=10$}
  \includegraphics[scale=0.35]{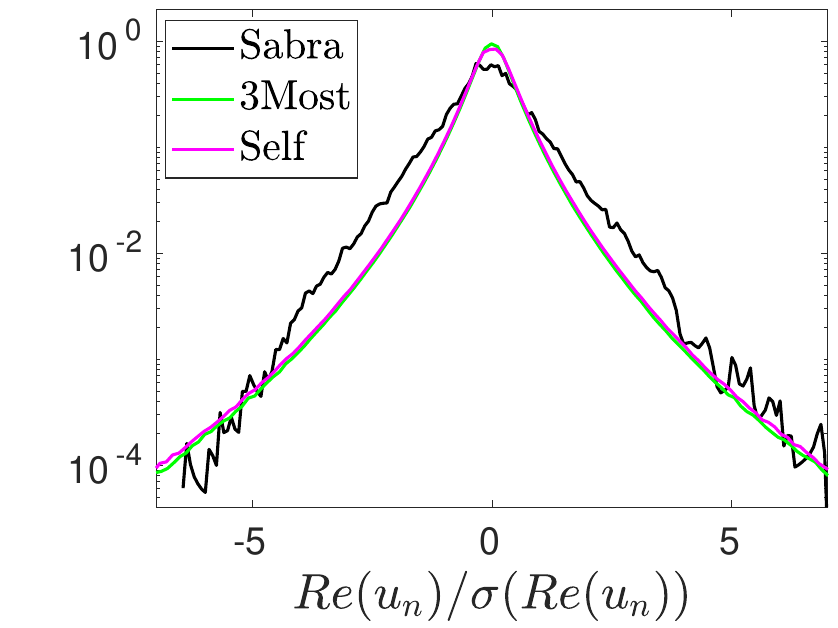}
  \label{fig:re139es}
\end{subfigure}%
\begin{subfigure}{.333333\textwidth}
  \centering
  \subcaption{$n=11$}
  \includegraphics[scale=0.35]{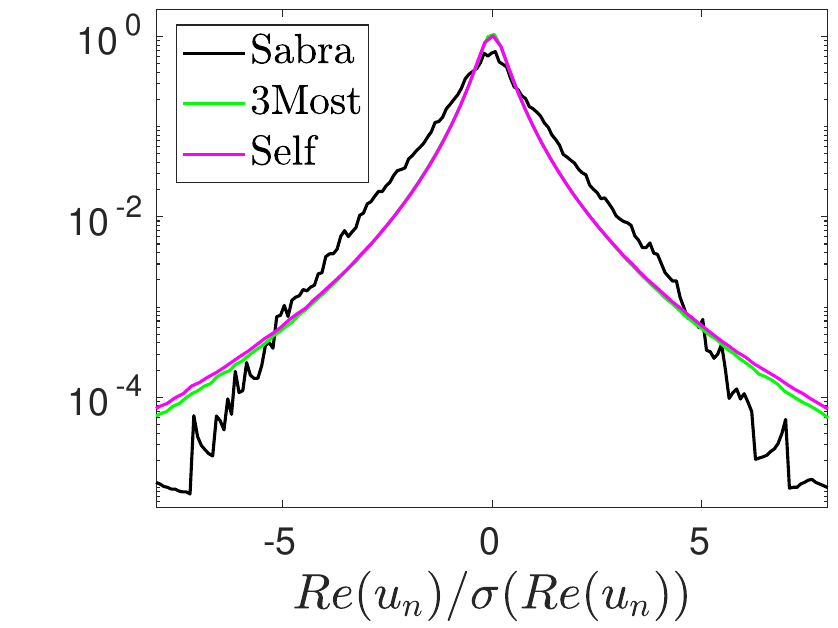}
  \label{fig:re149es}
\end{subfigure}%

\begin{subfigure}{.5\textwidth}
  \centering
  \subcaption{Self}
  \includegraphics[scale=0.56]{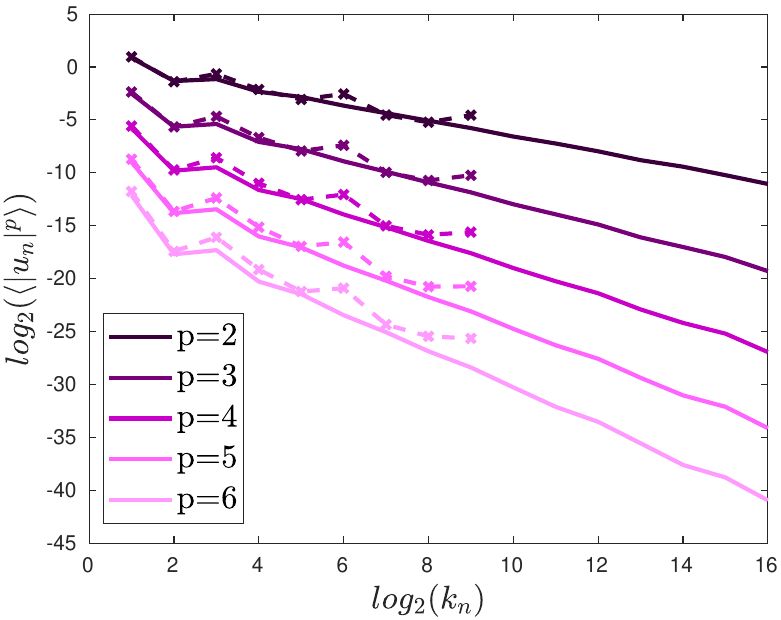}
  \label{fig:mjoint9es}
\end{subfigure}%
\begin{subfigure}{.5\textwidth}
  \centering
  \subcaption{3Most}
  \includegraphics[scale=0.56]{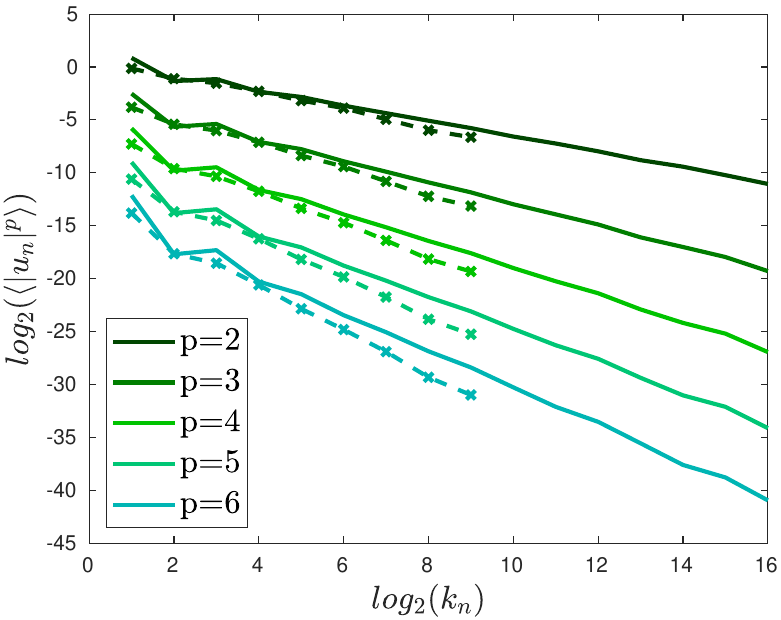}
  \label{fig:pjoint9es}
\end{subfigure}%

\begin{subfigure}{.333\textwidth}
  \centering
  \subcaption{$n=4$}
  \includegraphics[scale=0.35]{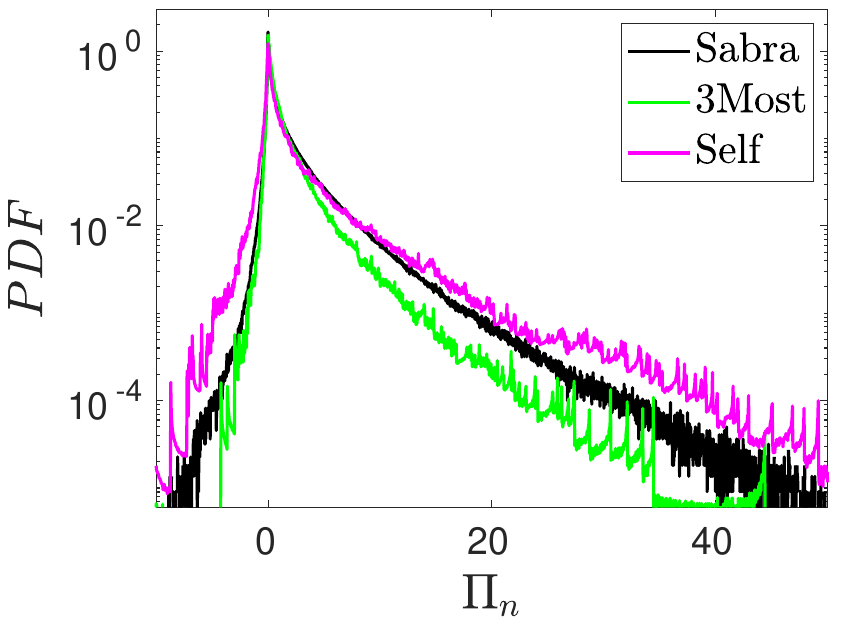}
  \label{fig:flx99es}
\end{subfigure}%
\begin{subfigure}{.333\textwidth}
  \centering
  \subcaption{$n=8$}
  \includegraphics[scale=0.35]{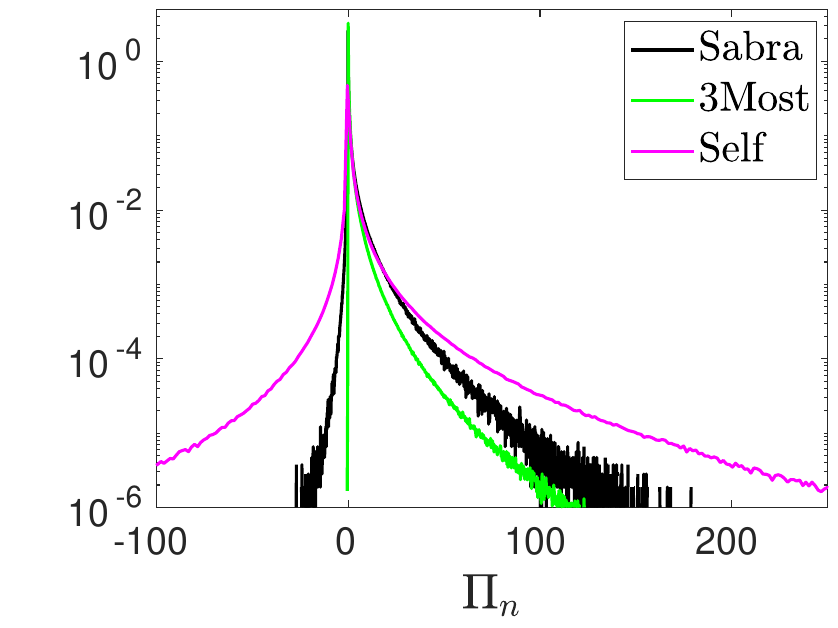}
  \label{fig:flx119es}
\end{subfigure}%
\begin{subfigure}{.333\textwidth}
  \centering
  \subcaption{$n=9$}
  \includegraphics[scale=0.35]{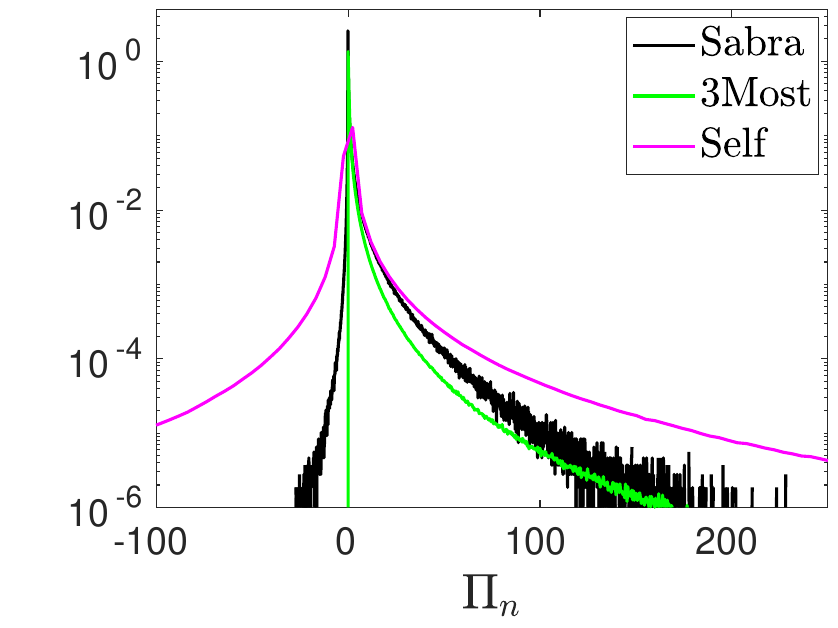}
  \label{fig:flx129es}
\end{subfigure}%
\caption{Cut-off shell is $s=9$. Figures \ref{fig:re129es}---\ref{fig:re149es} present normalized PDFs of real parts for the closure with time conditioning to the three closes shells (3Most) and the closure with time conditioning to the closure variables (Self). Figures \ref{fig:mjoint9es} and \ref{fig:pjoint9es} show moments of orders 2 up to 6 and figures \ref{fig:flx99es}---\ref{fig:flx12} show energy flux PDFs.}
\label{fig:3eself9}
\end{figure}

The crucial achievement in the formulation of our approach is its universality with respect to the cutoff scale $s$, because the statistics of the rescaled variables are universal through the inertial interval. Once we change the cut-off shell $s$ we just need to change the reference shell $m$ accordingly. We refer to figure \ref{fig:scaleinv} to recall that the densities for $\mathcal{U}_0$ and $\mathcal{U}_1$ do not change when we change the reference shell. To better see this, we will use the density estimations used in the simulations with the cut-off $s=12$ to run another simulations with cut-off shell $s=9$. The results are presented in Fig. \ref{fig:3eself9}, which correspond to the history conditioned closure in equations \eqref{eqn:3most0}--\eqref{eqn:dist3most}, as well as the closure in equations \eqref{eqn:s1s2ems1s20}--\eqref{eqn:s1s2ems1s2}. Comparison of Figs. \ref{fig:3eself} and \ref{fig:3eself9} shows that the performance of these closures is indeed independent of the cutoff scale $s$.
One can see from Figs. \ref{fig:self12} and \ref{fig:mjoint9es} that the behavior of closures near the cutoff have similar details, e.g., the same bump in the graphs of the moments at shells 9 and 6 respectively. Such a detailed coincidence was related in \cite{mailybaev2023hidden} to the extended hidden scale invariance of closures in shell models.

In summary, we observe that all proposed closures lead to a satisfactory performance per se, but the addition of the extra information such as conditioning on extra variables and previous times, do not lead to a noticeable improvement: some of the characteristics improve while others get worse. We discuss now the possible causes for such a behavior and the ways how it can be improved.

\section{Discussion}\label{sec_disc}

In this work, we addressed the question of whether it is possible and how to formulate a small-scale closure for the shell model of turbulence in such a way that it accurately reproduces the dynamics at all resolved scales of motion. The problem of formulating this closure is related to two important properties of turbulent solutions: intermittency~\cite{frisch} and spontaneous stochasticity~\cite{Mailybaev_2016}. We argue that the proper closure that satisfies these properties must be probabilistic and time-correlated to prehistory. Working with the Sabra model, we apply a space-time rescaling that reveals the hidden scaling symmetry and, hence, restores universality of multi-scale and multi-time correlations. We then use a density estimation process to approximate the density of rescaled closure variables, and apply this approximation to (conditionally) sample new instances of data from this density.  Lastly, use these new instances to evolve reduced models in time. 

The performance of the closures tested numerically in this work varied, but in general managed to recover at least some statistics of Sabra model with reasonable accuracy. Even though we succeeded in including the modelling of phases in our closures, we still detect persistent difficulties in obtaining a tight fit of PDFs of real parts and energy flux through the final shells. However, the fact that these closures work on data alone, without any direct adjustment or learning towards phenomenological predictions, can be seen as a success. 

We note that the limited performance of our models can be related to the choice of the GMM density estimator, which has difficulties in reproducing the statistics at high phase-space dimensions. One could improve from GMM to any other machine learning tool that scales better as the dimension of the optimization problem grows \cite{lstm,lstmsabra}. This highlights one of the main contribution of this work, which is the clear and systematic framework of universal statistics that open new possibilities for employing machine learning tools in turbulence problems.  

This framework is developed by connecting the concept of the closure with the hidden scaling symmetry. Such universal closures allow a systematic use of the information at the resolved scales, both at present and previous times (conditioning on the prehistory), which is not possible in the standard formulation due to the intermittency~\cite{Biferale_2017}. We stress that the strategy we used is not limited to shell models, but can potentially be generalized to the Navier--Stokes turbulence \cite{mailybaev2022hidden,Mailybaev2022HiddenSI}.

\appendix
\section{List of all closures considered}\label{app:all}
A wide variety of combinations between how many and in which shells we could conditionally sample the closure variables have been tested. Table \ref{tab:allclos} shows all closures tested, with the ones reported in section \ref{sec:results} in bold letters.  The first line on the table refers to Kolmogorov's closure \cite{Biferale_2017}. It is the deterministic closure given by the following expressions
\begin{gather}
    \mathcal{U}_{0} = |\mathcal{U}_{-1}|\lambda^{-1/3} e^{i(\frac{\pi}{2} + \alpha_{-1} + \alpha_{-2})}, \label{eqn:U0kolm} \\ 
    \mathcal{U}_{1} = |\mathcal{U}_{0}|\lambda^{-1/3} e^{i(\frac{\pi}{2} + \alpha_{0} + \alpha_{-1})} \label{eqn:U1kolm}. %= |\mathcal{U}_{-1}|\lambda^{-2/3} e^{i(\pi + 2\alpha_{-1} + \alpha_{-2})} 
\end{gather} \label{eqn:kolmnew}
This closure is also used to cover the initial time steps in simulations that require time conditioning. For more details about the performance of closured from Table \ref{tab:allclos} we refer to the PhD thesis \cite{memyselfandi}.
Table \ref{tab:theones} shows the closures reported in sections \ref{sec:results} and \ref{sec:withcond}, to clear the fact that they can be grouped in more than one way, referring to whether they include time conditioning or not, or whether they include phase modelling or not.

\begin{table}[]
\centering
\begin{tabular}{l|ccccc}
 &
  $|\mathcal{U}_0|$ &
  $|\mathcal{U}_1|$ &
  $\Delta_0$ &
  $\Delta_1$ &
  Conditioning \\ \hline
Kolmogorov &
  $|\mathcal{U}_{-1}|\lambda^{-1/3}$ &
  $|\mathcal{U}_0|\lambda^{-1/3}$ &
  $\pi/2$ &
  $\pi/2$ &
  $\times$ \\ \hline
Half closure &
  $2^{z_0}$ &
  $|\mathcal{U}_0|\lambda^{-1/3}$ &
  $\pi/2$ &
  $\pi/2$ &
  $\times$ \\ \hline
\textbf{Joint} &
  $2^{z_0}$ &
  $2^{z_1}$ &
  $\pi/2$ &
  $\pi/2$ &
  $\times$ \\ \hline
Simple cond &
  $2^{z_0}$ &
  $|\mathcal{U}_0|\lambda^{-1/3}$ &
  $\pi/2$ &
  $\pi/2$ &
  $|\mathcal{U}_{-1}|$ at $\tau - \Delta \tau$ \\ \hline
Joint cond &
  $2^{z_0}$ &
  $2^{z_1}$ &
  $\pi/2$ &
  $\pi/2$ &
  $|\mathcal{U}_{-1}|$ at $\tau - \Delta \tau$ \\ \hline
\textbf{3-Most} &
  $2^{z_0}$ &
  $2^{z_1}$ &
  $\pi/2$ &
  $\pi/2$ &
  $|\mathcal{U}_{-3}|,|\mathcal{U}_{-2}|,|\mathcal{U}_{-1}|$ at $\tau - \Delta \tau$ \\ \hline
\textbf{3-Most 9} &
  $2^{z_0}$ &
  $2^{z_1}$ &
  $\pi/2$ &
  $\pi/2$ &
  $|\mathcal{U}_{-3}|,|\mathcal{U}_{-2}|,|\mathcal{U}_{-1}|$ at $\tau - \Delta \tau$ \\ \hline
Long &
  $2^{z_0}$ &
  $2^{z_1}$ &
  $\pi/2$ &
  $\pi/2$ &
  $|\mathcal{U}_{-s}|,\dots,|\mathcal{U}_{-1}|$ at $\tau - \Delta \tau$ \\ \hline
\textbf{Joint phases} &
  $2^{z_0}$ &
  $2^{z_2}$ &
  $z_1$ &
  $z_3$ &
  $\times$ \\ \hline
Joint cond phases &
  $2^{z_0}$ &
  $2^{z_2}$ &
  $z_1$ &
  $z_3$ &
  $|\mathcal{U}_{-1}|,\Delta{-1}$ at $\tau - \Delta \tau$ \\ \hline
\textbf{Self} &
  $2^{z_0}$ &
  $2^{z_2}$ &
  $z_1$ &
  $z_3$ &
  $|\mathcal{U}_{0}|,|\mathcal{U}_{1}|,\Delta_{0},\Delta_{1}$ at $\tau - \Delta \tau$ \\ \hline
\textbf{Self 9} &
  $2^{z_0}$ &
  $2^{z_2}$ &
  $z_1$ &
  $z_3$ &
  $|\mathcal{U}_{0}|,|\mathcal{U}_{1}|,\Delta_{0},\Delta_{1}$ at $\tau - \Delta \tau$ \\ \hline
Global &
  $2^{z_0}$ &
  $2^{z_2}$ &
  $z_1$ &
  $z_3$ &
  \begin{tabular}[c]{@{}c@{}}$|\mathcal{U}_{-2}|,|\mathcal{U}_{-1}|,\Delta_{-2},\Delta_{-1}$ at $\tau$ \\ $|\mathcal{U}_{-2}|,|\mathcal{U}_{-1}|,|\mathcal{U}_{0}|,|\mathcal{U}_{1}|,\Delta_{-2},\Delta_{-1},\Delta_{0},\Delta_{1}$ at $\tau - \Delta \tau$\end{tabular} \\ \hline
2 Times &
  $2^{z_0}$ &
  $2^{z_2}$ &
  $z_1$ &
  $z_3$ &
  \begin{tabular}[c]{@{}c@{}}$|\mathcal{U}_{-2}|,|\mathcal{U}_{-1}|,\Delta_{-2},\Delta_{-1}$ at $\tau$ \\ $|\mathcal{U}_{-2}|,|\mathcal{U}_{-1}|,|\mathcal{U}_{0}|,|\mathcal{U}_{1}|,\Delta_{-2},\Delta_{-1},\Delta_{0},\Delta_{1}$ at $\tau - \Delta \tau$\\ $|\mathcal{U}_{-2}|,|\mathcal{U}_{-1}|,|\mathcal{U}_{0}|,|\mathcal{U}_{1}|,\Delta_{-2},\Delta_{-1},\Delta_{0},\Delta_{1}$ at $\tau - 2\Delta \tau$\end{tabular}
\end{tabular}
\caption{All closures written for this work, with the ones in bold being the ones presented here.}
\label{tab:allclos}
\end{table}

\begin{table}[]
\centering
\begin{tabular}{l|cc}
                & Includes phases      & Only modules \\ \hline
Conditioning    & Self, Self 9         & 3-Most, 3-Most 9           \\
No conditioning & Joint phases       & Joint                    
\end{tabular}
\caption{The closures we presented in this paper}
\label{tab:theones}
\end{table}

\section{The rescaled system}\label{app:proof}

Here we derive the equations of motion in terms of intrinsic time $\tau$ for the rescaled variables $\mathcal{U}_N$. From the definition of $\tau$ in equation \eqref{eqn:taut}, we have
\begin{eqnarray}\label{eqn:deriving1}
    \frac{d\mathcal{U}_N}{d\tau} = \frac{d\mathcal{U}_N}{dt} \frac{dt}{d\tau}.
\end{eqnarray}
The first factor can be computed deriving equation \eqref{eqn:UN} with respect to $t$, finding
\begin{align}
    \frac{d\mathcal{U}_N}{dt} = \frac{d T_m}{dt}k_m u_{N+m} + T_m k_m \frac{du_{N+m}}{dt}.
\end{align}
and the second can be computed rewriting equation \eqref{eqn:taut} as
\begin{align}
    \frac{dt}{d\tau} = T_m. \label{eqn:change}
\end{align}
Equation \eqref{eqn:deriving1} then becomes
\begin{align}\label{eqn:system1}
    \frac{d\mathcal{U}_N}{d\tau} = \left( \frac{d T_m}{dt}k_m u_{N+m} + T_m k_m \frac{du_{N+m}}{dt} \right) T_m.
\end{align}

We compute the derivative of $T_m$ with respect to $t$ as
\begin{align}
    \frac{dT_m}{dt} = -\frac{1}{2} \left( k_0^2 U^2 + \sum_{n<m} k_n^2 |u_n|^2 \right) ^{-3/2}  \sum_{n<m} k_n^2 2\operatorname{Re} \left( u_n^*\frac{du_n}{dt} \right).
\end{align}
Using equations \eqref{eqn:Tm} and \eqref{eqn:sabra} in the above expression, we find
\begin{align}
    \frac{dT_m}{dt} = -T_m^3 \sum_{n<m} k_n^2 \operatorname{Re}( u_n^* [i(k_{n+1}u_{n+2}u_{n+1}^* -\frac{1}{2}k_n u_{n+1}u_{n-1}^* + \frac{1}{2}k_{n-1}u_{n-1}u_{n-2}) - \nu k_n^2u_n + f_n  ] ),
\end{align}

\noindent which is the same as
\begin{align}
    \frac{dT_m}{dt} = - \sum_{n<m} \operatorname{Re} \Big{(} i( k_n T_m u_n^* & k_n T_m u_{n+2}  k_{n+1}T_m u_{n+1}^* \nonumber \\
    - \frac{1}{2}k_n T_m u_n^* & k_n T_m u_{n+1}  k_n T_m u_{n-1}^* \nonumber \\
    + \frac{1}{2} k_n T_m u_n^* & k_{n-1} T_m u_{n-1}  k_n T_m u_{n-2}) \nonumber \\
    - \nu k_n^4 T_m^3 u_n^* u_n &+k_n^2 T_m^3 u_n^* f_n \Big{)}.
\end{align}

A direct manipulation, noting that $k_n = k_{n-m} k_m$, $k_1=2$, $k_{-1}=1/2$, rewriting the indexes as $N=n-m$ and recalling that $n<m$ is the same as $N<0$, yields
% Now note that we need some $k_m$ to appear in each parcel in order to use the change of variables defined in equation \eqref{eqn:UN}, so we write $k_n = k_{n-m} k_m$ and that leads to
% \begin{align}
%     \frac{dT_m}{dt} = - \sum_{n<m} \operatorname{Re} \Big{(} i(k_{n-m}^2 k_{n+1-m} \cdot k_m T_m u_n^* &\cdot k_m T_m u_{n+2} \cdot k_m T_m u_{n+1}^* \nonumber \\
%     - \frac{1}{2}k_{n-m}^3 \cdot k_m T_m u_n^* &\cdot k_m T_m u_{n+1} \cdot k_m T_m u_{n-1}^*\nonumber \\
%     + \frac{1}{2} k_{n-m}^2 k_{n-1-m} \cdot k_n T_m u_n^* &\cdot k_{n-1} T_m u_{n-1} \cdot k_n T_m u_{n-2}) \nonumber \\
%     - \nu k_n^2 k_{n-m}^2 k_m^2 T_m^3 u_n^* u_n &+ k_nk_{n-m}k_m T_m^3 u_n^* f_n \Big{)}.
% \end{align}
% We can now go ahead and do the change of variables from equation \eqref{eqn:UN}, finding
% \begin{align}
%     \frac{dT_m}{dt} = - \sum_{n<m} \operatorname{Re} \Big{(} i(k_{n-m}^3k_1  \cdot \mathcal{U}_{n-m}^* &\cdot \mathcal{U}_{n+2-m} \cdot \mathcal{U}_{n+1-m}^* \nonumber \\
%     - \frac{1}{2}k_{n-m}^3 \cdot \mathcal{U}_{n-m}^* &\cdot \mathcal{U}_{n+1-m} \cdot \mathcal{U}_{n-1-m}^* \nonumber \\
%     + \frac{1}{2} k_{n-m}^3 k_{-1} \cdot \mathcal{U}_{n-m}^* &\cdot \mathcal{U}_{n-1-m} \cdot \mathcal{U}_{n-2-m}) \nonumber \\
%     - \nu k_n^2 k_{n-m}^2 T_m |\mathcal{U}_{n-m}|^2 &+ k_n k_{n-m} T_m^2 \mathcal{U}_{n-m}^* f_n \Big{)}.
% \end{align}
% Now note that $k_1=2$, $k_{-1}=1/2$ and let us re-write the indices, calling $N=n-m$ and seeing that $n<m$ is the same as $N<0$. We then have
\begin{align}
    \frac{dT_m}{dt} = - \sum_{N<0} \operatorname{Re} \Big{(} i(k_{N}^3 2  \mathcal{U}_{N}^* & \mathcal{U}_{N+2} \mathcal{U}_{N+1}^* \nonumber \\
    - \frac{1}{2}k_{N}^3  \mathcal{U}_{N}^* & \mathcal{U}_{N+1}  \mathcal{U}_{N-1}^* 
    + \frac{1}{4} k_{N}^3  \mathcal{U}_{N}^*  \mathcal{U}_{N-1}  \mathcal{U}_{N-2}) \nonumber \\
    - \nu k_{N+m}^2 k_{N}^2 T_m |\mathcal{U}_{N}|^2 &+ k_{N+m} k_{N} T_m^2 \mathcal{U}_{N}^* f_{N+m} \Big{)},
\end{align}
which can be further reduced to the form
\begin{align}
    \frac{dT_m}{dt} =  \sum_{N<0} k_{N}^3 \operatorname{Im} \Big{(}  2 \mathcal{U}_{N}^*  \mathcal{U}_{N+1}^* \mathcal{U}_{N+2} - \frac{1}{2}& \mathcal{U}_{N-1}^* \mathcal{U}_{N}^* \mathcal{U}_{N+1} - \frac{1}{4} \mathcal{U}_{N-1}^*  \mathcal{U}_{N}  \mathcal{U}_{N-2}^* \Big{)} \nonumber \\
    + \nu T_m \sum_{N<0} k_{N+m}^2 k_{N}^2 |\mathcal{U}_{N}|^2 &- T_m^2 \sum_{N<0}k_{N+m} k_{N} \operatorname{Re}\Big{(} \mathcal{U}_{N}^* f_{N+m} \Big{)}.
\end{align}
Splitting the sum above in several pieces, we write
\begin{gather} 
    \frac{dT_m}{dt} = \xi + \xi_{\nu} + \xi_f = \xi_{total}, \tag{\ref{eqn:dtmdt}} \\
    \xi = \sum_{N<0} k_{N}^3 \operatorname{Im} \Big{(}  2 \mathcal{U}_{N}^*  \mathcal{U}_{N+1}^* \mathcal{U}_{N+2} - \frac{1}{2} \mathcal{U}_{N-1}^* \mathcal{U}_{N}^* \mathcal{U}_{N+1} - \frac{1}{4} \mathcal{U}_{N-1}^*  \mathcal{U}_{N}  \mathcal{U}_{N-2}^* \Big{)}, \tag{\ref{eqn:xi}}\\
    \xi_{\nu} = \nu T_m k_m^2 \sum_{N<0}  k_{N}^4 |\mathcal{U}_{N}|^2 , \tag{\ref{eqn:xinu}}\\
    \xi_f = - T_m^2 \sum_{N<0}k_{N+m} k_{N} \operatorname{Re}\Big{(} \mathcal{U}_{N}^* f_{N+m} \Big{)}.  \tag{\ref{eqn:xif}}
\end{gather}

% Combining (\ref{eqn:dtmdt}) with \eqref{eqn:change}, we have

% \begin{equation}
% \frac{dT_m}{d\tau} = (\xi + \xi_{\nu} + \xi_f)T_m. \label{eqn:20}
% \end{equation}

We can now go back to computing the main system, plugging \cref{eqn:xi,eqn:xinu,eqn:xif,eqn:dtmdt} back into \eqref{eqn:system1}. This yields
\begin{align}\label{eqn:system2}
    \frac{d\mathcal{U}_N}{d\tau} = \left( \xi_{total} k_m u_{N+m} + T_m k_m \frac{du_{N+m}}{dt} \right) T_m.
\end{align}

We then use equation \eqref{eqn:sabra}, with $b=-1/2$ and $c=1/2$, to write
\begin{align}\label{eqn:system3}
    \frac{d\mathcal{U}_N}{d\tau} = \xi_{total}T_m k_m u_{N+m} + T_m^2 k_m& \Big{(}  i(k_{N+m+1}u_{N+m+2}u_{N+m+1}^* \nonumber \\
    &-\frac{1}{2}k_{N+m} u_{N+m+1}u_{N+m-1}^* \nonumber \\
    &+ \frac{1}{2}k_{N+m-1}u_{N+m-1}u_{N+m-2}) - \nu k_{N+m}^2u_{N+m} + f_{N+m} \Big{)},
\end{align}

\noindent which becomes, doing the change of variables from \eqref{eqn:UN} again,
% \begin{align}\label{eqn:system4}
%     \frac{d\mathcal{U}_N}{d\tau} = \xi_{total}T_m k_m u_{N+m} + & i(k_{N+1}T_m k_mu_{N+m+2}T_m k_mu_{N+m+1}^* \nonumber \\
%     &-\frac{1}{2}k_{N} T_m k_mu_{N+m+1}T_m k_mu_{N+m-1}^* \nonumber \\
%     &+ \frac{1}{2}k_{N-1}T_m k_mu_{N+m-1}T_m k_mu_{N+m-2}) \nonumber \\
%     &- \nu k_{N+m}^2T_m^2 k_mu_{N+m} + T_m^2 k_mf_{N+m}.
% \end{align}
% Now let us do the change of variables from \eqref{eqn:UN} again, to find
\begin{align}\label{eqn:system5}
    \frac{d\mathcal{U}_N}{d\tau} = \xi_{total}\mathcal{U}_{N} + & i(k_{N+1}\mathcal{U}_{N+2}\mathcal{U}_{N+1}^* \nonumber \\
    &-\frac{1}{2}k_{N} \mathcal{U}_{N+1}\mathcal{U}_{N-1}^* \nonumber \\
    &+ \frac{1}{2}k_{N-1}\mathcal{U}_{N-1}\mathcal{U}_{N-2}) - \nu k_{N+m}^2T_m \mathcal{U}_{N} + T_m^2 k_m f_{N+m}.
\end{align}
Reorganizing the terms, we have
\begin{align}
    \frac{d\mathcal{U}_N}{d\tau} = i(k_{N+1}\mathcal{U}_{N+2}\mathcal{U}_{N+1}^* -\frac{1}{2}k_{N} \mathcal{U}_{N+1}\mathcal{U}_{N-1}^* + \frac{1}{2}k_{N-1}\mathcal{U}_{N-1}\mathcal{U}_{N-2}) \nonumber \\
    +\left( \xi_{total} - \nu k_{N+m}^2T_m \right) \mathcal{U}_{N} + T_m^2 k_m f_{N+m}. \tag{\ref{eqn:system6}}
\end{align}

% Now let us look specifically at the term $( \xi_{total} - \nu k_{N+m}^2T_m) \mathcal{U}_{N}$ and use equation \eqref{eqn:xinu} to write the viscous term.

% \begin{align}
%     \left( \xi_{total} - \nu k_{N+m}^2T_m \right) \mathcal{U}_{N} = \left( \xi + \xi_{\nu} + \xi_f - \nu k_{N+m}^2T_m\right) \mathcal{U}_N \nonumber \\
%      = \left( \xi + \xi_f  +\nu k_m^2T_m\sum_{N<0}\left(k_N^4|\mathcal{U}_N|^2\right) - \nu  k_{N+m}^2T_m\right) \mathcal{U}_N \nonumber \\
%      =\left( \xi + \xi_f + \nu k_m^2T_m\big( -k_N^2 + \sum_{N<0} k_N^4|\mathcal{U}_N|^2\big)\right) \mathcal{U}_N \label{eqn:viscous}
% \end{align}

% We can then plug equation \eqref{eqn:viscous} back into equation \eqref{eqn:system3} and we will have

% \begin{align}\label{eqn:systemfull}
%     \frac{d\mathcal{U}_N}{d\tau} = i(k_{N+1}\mathcal{U}_{N+2}\mathcal{U}_{N+1}^* -\frac{1}{2}k_{N} \mathcal{U}_{N+1}\mathcal{U}_{N-1}^* + \frac{1}{2}k_{N-1}\mathcal{U}_{N-1}\mathcal{U}_{N-2}) \nonumber \\
%     +\left( \xi + \xi_f \right) \mathcal{U}_N + \nu k_m^2T_m\big( -k_N^2 + \sum_{J<0} k_J^4|\mathcal{U}_J|^2\big) \mathcal{U}_N + T_m^2 k_m f_{N+m}
% \end{align}
In order to write this system with no dependence on $t$, we must also write $T_m$ as a function of $\mathcal{U}$ instead of $u$. Recall the expression for $T_m$ in equation \eqref{eqn:Tm}, expand the sum and apply the change of variables in equation \eqref{eqn:UN}
% \begin{align}
% T_m = \left(k_0^2 U^2 + \sum_{n <m} k_n^2|u_n|^2 \right)^{-1/2}. \nonumber  
% \end{align}
%Looking at the expanded sum and applying the change of variables, we see that
% \begin{align}
%     T_m = \left(k_0^2 U^2 +  k_1^2|u_1|^2 + k_2^2|u_2|^2 + \cdots + k_{m-1}^2|u_{m-1}|^2 \right)^{-1/2}.
% \end{align}
% Now doing the change of variables defined in \eqref{eqn:UN}, 
\begin{align}
    T_m = \left(k_0^2 U^2 +  \frac{k_1^2}{k_m^2 T_m^2} |\mathcal{U}_{1-m}|^2 + \frac{k_2^2}{k_m^2 T_m^2} |\mathcal{U}_{2-m}|^2 + \cdots + \frac{k_{m-1}^2}{k_m^2 T_m^2} |\mathcal{U}_{-1}|^2 \right)^{-1/2}.
\end{align}

% Dividing by $k_m^2$ where indicated,
% \begin{align}
%     T_m = \left(k_0^2 U^2 +  \frac{k_{1-m}^2}{T_m^2} |\mathcal{U}_{1-m}|^2 + \frac{k_{2-m}^2}{T_m^2} |\mathcal{U}_{2-m}|^2 + \cdots + \frac{k_{-1}^2}{ T_m^2} |\mathcal{U}_{-1}|^2 \right)^{-1/2}.
% \end{align}
Regrouping the sum,
\begin{align}
    \frac{1}{T_m^2} = k_0^2 U^2 +  \frac{1}{T_m^2}\sum_{N<0}k_N^2|\mathcal{U}_N|^2.
\end{align}

It then follows that
\begin{gather}
    % 1-k_0^2U^2T_m^2 = \sum_{N<0}k_N^2|\mathcal{U}_N|^2, \\
    % T_m^2 = \frac{1}{k_0^2U^2}\left( 1- \sum_{N<0}k_N^2|\mathcal{U}_N|^2 \right), \\
    T_m = \frac{1}{k_0 U}\left( 1- \sum_{N<0}k_N^2|\mathcal{U}_N|^2 \right)^{1/2}. \tag{\ref{eqn:Tm_tau}}
\end{gather}

% To summarize, we have the rescaled viscous, forced system given by
% \begin{align}
%     \frac{d\mathcal{U}_N}{d\tau} = i(k_{N+1}\mathcal{U}_{N+2}\mathcal{U}_{N+1}^* -\frac{1}{2}k_{N} \mathcal{U}_{N+1}\mathcal{U}_{N-1}^* + \frac{1}{2}k_{N-1}\mathcal{U}_{N-1}\mathcal{U}_{N-2}) \nonumber \\
%     +\left( \xi_{total} - \nu k_{N+m}^2T_m \right) \mathcal{U}_{N} + T_m^2 k_m f_{N+m}, \tag{\ref{eqn:system6}}
% \end{align}

% \noindent where
% \begin{gather} 
%     \xi_{total} = \xi + \xi_{\nu} + \xi_f,  \tag{\ref{eqn:dtmdt}} \\
%     \xi = \sum_{N<0} k_{N}^3 \operatorname{Im} \Big{(}  2 \mathcal{U}_{N}^*  \mathcal{U}_{N+1}^* \mathcal{U}_{N+2} - \frac{1}{2} \mathcal{U}_{N-1}^* \mathcal{U}_{N}^* \mathcal{U}_{N+1} - \frac{1}{4} \mathcal{U}_{N-1}^*  \mathcal{U}_{N}  \mathcal{U}_{N-2}^* \Big{)}, \tag{\ref{eqn:xi}}\\
%     \xi_{\nu} = \nu T_m k_m^2 \sum_{N<0}  k_{N}^4 |\mathcal{U}_{N}|^2 , \tag{\ref{eqn:xinu}}\\
%     \xi_f = - T_m^2 \sum_{N<0}k_{N+m} k_{N} \operatorname{Re}\Big{(} \mathcal{U}_{N}^* f_{N+m} \Big{)},  \tag{\ref{eqn:xif}} \\ 
%     T_m = \frac{1}{k_0 U}\left( 1- \sum_{N<0}k_N^2|\mathcal{U}_N|^2 \right)^{1/2}, \tag{\ref{eqn:Tm_tau}}
% \end{gather}

% \noindent as we wanted.

\bibliographystyle{siam}
\bibliography{sample}

\end{document}